# THE THEORY OF ARTIFICIAL IMMUTABILITY: PROTECTING ALGORITHMIC GROUPS UNDER ANTI-DISCRIMINATION LAW

*Sandra Wachter*[1,2]

## ABSTRACT

Artificial Intelligence (AI) and machine learning algorithms are increasingly used to make important decisions about people. Decisions taken on the basis of socially defined groups can have harmful consequences, creating unequal, discriminatory, and unfair outcomes on the basis of irrelevant or unacceptable differences. Equality and anti-discrimination laws aim to protect against these types of harms.

While issues of AI bias and proxy discrimination are well explored, less focus has been paid to the harms created by profiling based on groups that do not map to or correlate with legally protected groups such as sex or ethnicity. Groups like dog owners, sad teens, video gamers, single parents, gamblers, or the poor are routinely used to allocate resources and make decisions such as which advertisement to show, price to offer, or public service to fund. AI also creates seemingly incomprehensible groups defined by parameters that defy human understanding such as pixels in a picture, clicking behavior, electronic signals, or web traffic. These algorithmic groups feed into important automated decisions, such as loan or job applications, that significantly impact people's lives.

A technology that groups people in unprecedented ways and makes decisions about them naturally raises a question: are our existing equality laws, at their core, fit for purpose to protect against emergent, AI-driven inequality?

This paper examines the legal status of algorithmic groups in North American and European anti-discrimination doctrine, law, and jurisprudence. I propose a new theory of harm to close the gap between legal doctrine and emergent forms of algorithmic discrimination. Algorithmic groups do not currently enjoy legal protection unless they can be mapped onto an existing protected group. Such linkage is rare in practice. In response, this paper examines three possible pathways to expand the scope of anti-discrimination law to include algorithmic groups.

The first possibility is to show that algorithmic groups meet the law's requirements to be considered a protected ground. A novel taxonomy of characteristics that make groups worthy of protection is developed, covering (1) immutability and choice, (2) relevance, arbitrariness, and merit, (3) historical oppression, stigma, and structural disadvantage, and (4) social saliency. Algorithmic groups typically do not meet these requirements in practice.

---

[1] Corresponding author: Sandra Wachter, Associate Professor and Senior Research Fellow, Oxford Internet Institute, University of Oxford, 1 St. Giles, Oxford, OX1 3JS, UK. E-mail: sandra.wachter@oii.ox.ac.uk

[2] The work of the Governance of Emerging Technologies research program at the Oxford Internet Institute has been supported by the British Academy Postdoctoral Fellowships (grant nr PF2\180114 and grant nr PF\170151), the Luminate Group, the Miami Foundation, the UK Department of Health and Social Care (via the AI Lab at NHSx), the Sloan Foundation (grant nr G-2021-16779), and the Wellcome Trust (grant nr 223765/Z/21/Z). The author is indebted to Brent Mittelstadt, Chris Russell, Johann Laux and Rory Gillis for their incredibly detailed and helpful feedback that greatly improved the quality of the manuscript.

The second possibility is to examine the law's theoretical account of the "moral wrongness" of discrimination. I will show that algorithmic groups do not invoke the same moral wrongfulness that the law wants to prevent.

The third possibility is to find an opening in the fundamental doctrines and remit of anti-discrimination law. Many scholars agree that the law aims to eliminate differences between certain groups, or to create a 'level playing field'. This approach creates a problem for algorithmic groups because it may be impossible to establish the existence of oppression, or to show that certain groups receive better treatment than others.

A new theory of harm is required if algorithmic groups ought to be protected under the law. Different theories of equality share common ground in their commitment to protect basic rights, liberties, freedoms, and access to goods and services. This common ground provides an opportunity to bring algorithmic groups within the scope of the law. The harms and interests facing these algorithmic groups are the same as those targeted by the law; only the mode, perpetrator, and process of bringing about the harms are different.

The paper closes with two contributions to drive legal reform and judicial reinterpretation. Firstly, I propose four key elements of a good decision criteria reflecting the law's fundamental aims: stability, transparency, empirical coherence, and ethical and normative acceptability. AI makes these criteria effectively impossible to meet in practice.

Secondly, I propose a new theory of harm, the "theory of artificial immutability," that aims to bring AI groups within the scope of the law. My theory describes how algorithmic groups act as *de facto* immutable characteristics in practice. I propose five sources of artificial immutability in AI: opacity, vagueness, instability, involuntariness and invisibility, and a lack of social concept. Each of these erodes the key elements of good decision criteria.

To remedy this, greater emphasis needs to be placed on whether people have control over decision criteria and whether they are able to achieve important goals and steer their path in life. I conclude with reflections on how the law can be reformed to account for artificial immutability, making use of a fruitful overlap with prior work on the "right to reasonable inferences."

# TABLE OF CONTENTS

## I.   INTRODUCTION AND BACKGROUND

Artificial intelligence (AI) and machine learning algorithms are increasingly used to make decisions about people and are widely deployed in both the private and the public sector. To make these decisions, AI creates groups by sifting through large amounts of data.

Groups are created based on small patterns and correlations in similar attributes, behaviors or preferences of individuals.[3] Risk stratification, credit scoring,[4] search and media filtration,[5] market segmentation, employment,[6] policing and criminal sentencing[7] are all areas already using group profiling to make important decisions.[8] AI is also used to decide who gets promoted, hired, fired,[9] who will be successful with their loan application,[10] who will be granted insurance, who gets medical treatment[11] and who gets admitted to university.[12]

---

[3] Brent Mittelstadt, *From Individual to Group Privacy in Big Data Analytics*, 30 PHILOSOPHY & TECHNOLOGY 475–494, 476 (2017); Mireille Hildebrandt, *Defining Profiling: A New Type of Knowledge?*, in PROFILING THE EUROPEAN CITIZEN 17–45, 17–45 (Mireille Hildebrandt & Serge Gutwirth eds., 2008), http://link.springer.com/chapter/10.1007/978-1-4020-6914-7_2 (last visited May 14, 2015); Bart W. Schermer, *The limits of privacy in automated profiling and data mining*, 27 COMPUTER LAW & SECURITY REVIEW 45–52, 45–52 (2011).

[4] Pauline T. Kim & Erik A. Hanson, *People Analytics and the Regulation of Information under the Fair Credit Reporting Act*, 61 ST. LOUIS U. L.J. 17 (2016).

[5] Jaron Harambam, Natali Helberger & Joris van Hoboken, *Democratizing algorithmic news recommenders: how to materialize voice in a technologically saturated media ecosystem*, 376 PHILOSOPHICAL TRANSACTIONS OF THE ROYAL SOCIETY A: MATHEMATICAL, PHYSICAL AND ENGINEERING SCIENCES 20180088 (2018); Stefan Bechtold & Catherine Tucker, *Trademarks, Triggers, and Online Search*, 11 JOURNAL OF EMPIRICAL LEGAL STUDIES 718–750 (2014); on online price setting, see Frederik Zuiderveen Borgesius, *Algorithmic Decision-Making, Price Discrimination, and European Non-discrimination Law*, EUROPEAN BUSINESS LAW REVIEW (FORTHCOMING) (2019); Ariel Ezrachi & Maurice E. Stucke, *The rise of behavioural discrimination*, EUROPEAN COMPETITION LAW REVIEW (2016).

[6] Pauline T Kim, *Data-Driven Discrimination at Work*, 58 81; Pauline T. Kim, *Manipulating opportunity*, 106 VIRGINIA LAW REVIEW 867–935 (2020).

[7] Alexandra Chouldechova, *Fair Prediction with Disparate Impact: A Study of Bias in Recidivism Prediction Instruments*, 5 BIG DATA 153–163 (2017); Marion Oswald & Alexander Babuta, *Data Analytics and Algorithmic Bias in Policing* (2019).

[8] For all of these examples and more, see Mittelstadt, *supra* note 3 at 476.

[9] JEREMIAS PRASSL, HUMANS AS A SERVICE: THE PROMISE AND PERILS OF WORK IN THE GIG ECONOMY (2018); Jeremias Prassl & Martin Risak, *Uber, taskrabbit, and co.: Platforms as employers-rethinking the legal analysis of crowdwork*, 37 COMP. LAB. L. & POL'Y J. 619 (2015); Vili Lehdonvirta, *Algorithms that divide and unite: delocalisation, identity and collective action in 'microwork'*, in SPACE, PLACE AND GLOBAL DIGITAL WORK 53–80 (2016).

[10] Talia B. Gillis, *False Dreams of Algorithmic Fairness: The Case of Credit Pricing*, SSRN JOURNAL (2020), https://www.ssrn.com/abstract=3571266 (last visited Jan 8, 2021).

[11] I. G. Cohen et al., *The Legal And Ethical Concerns That Arise From Using Complex Predictive Analytics In Health Care*, 33 HEALTH AFFAIRS 1139–1147 (2014); Sara Gerke, Timo Minssen & Glenn Cohen, *Ethical and legal challenges of artificial intelligence-driven healthcare*, ARTIFICIAL INTELLIGENCE IN HEALTHCARE 295–336 (2020); BARBARA PRAINSACK, PERSONALIZED MEDICINE: EMPOWERED PATIENTS IN THE 21ST CENTURY?: 7 (1st edition ed. 2017).

[12] Brent Mittelstadt et al., *The ethics of algorithms: Mapping the debate*, 3 BIG DATA & SOCIETY (2016), http://bds.sagepub.com/lookup/doi/10.1177/2053951716679679 (last

Undoubtedly this technology can bring about great benefits. AI can see patterns that the human eye cannot spot and can shed light on unknown correlations which can help to make more accurate predictions. It can group candidates in unexpected but useful ways and can make decision-making processes more efficient and coherent across a wide range of sectors such as credit, hiring insurance, education, and health.

At the same time AI also faces several legal and ethical problems. Discrimination is a persistent problem in algorithmic decision-making. Algorithms need to sift through large amounts of historical data in order to create profiles and classify cases for decisions. Unfortunately, very often these datasets are marked by historical disadvantage and the oppression of the past. Hiring algorithms, for example, can pick up biased hiring patterns and apply them to future employment decisions.[13] The potential of AI to perpetuate and exacerbate traditional forms of bias and inequality based on legally protected grounds is already well explored in the literature. Profiling groups have, for example, been shown to be proxies of protected groups under anti-discrimination law.[14]

The challenges of bias, fairness, and discrimination in AI cannot, however, be traced solely back to historical patterns of oppression and inequality. Questions of bias and inequality against groups that are not simply proxies of currently protected groups are less well explored. AI creates and works with new types of groups (for purposes of profiling or decision-making) that need not map onto or correlate with traditionally protected groups, and thus do not automatically receive protection under anti-discrimination law.

There are two novel types of groups to consider in this context. I refer to these collectively as "algorithmic groups" in recognition of their creation (or usage) by AI systems to make important decisions. Examples already used for purposes such as advertising and personalized pricing include

---

visited Dec 15, 2016); Ofqual's A-level algorithm: why did it fail to make the grade?, THE GUARDIAN (2020), http://www.theguardian.com/education/2020/aug/21/ofqual-exams-algorithm-why-did-it-fail-make-grade-a-levels (last visited Jan 17, 2021).

[13] Reuters, *Amazon ditched AI recruiting tool that favored men for technical jobs*, THE GUARDIAN, October 10, 2018, https://www.theguardian.com/technology/2018/oct/10/amazon-hiring-ai-gender-bias-recruiting-engine (last visited Mar 2, 2020); for more on how to counter bias in hiring, see PAULINE KIM, *Race-Aware Algorithms: Fairness, Nondiscrimination and Affirmative Action*, (2022), https://papers.ssrn.com/abstract=4018414 (last visited Feb 7, 2022).

[14] Latanya Sweeney, *Discrimination in online ad delivery*, 11 QUEUE 1–19 (2013); Ruha Benjamin, *Race after technology: Abolitionist tools for the new jim code*, SOCIAL FORCES (2019); VIRGINIA EUBANKS, AUTOMATING INEQUALITY: HOW HIGH-TECH TOOLS PROFILE, POLICE, AND PUNISH THE POOR (2018); CATHY O'NEIL, WEAPONS OF MATH DESTRUCTION: HOW BIG DATA INCREASES INEQUALITY AND THREATENS DEMOCRACY (2017); CAROLINE CRIADO PEREZ, INVISIBLE WOMEN: EXPOSING DATA BIAS IN A WORLD DESIGNED FOR MEN (2019).

"dog owners,"[15] "sad teens,"[16] "video gamers,"[17] "single parents,"[18] "gamblers,"[19] and "poor people."[20] The first type, which I will refer to as "non-protected," are groups defined by human-comprehensible characteristics and definition rules that do not currently enjoy protection under equality and anti-discrimination law.

The second, which I refer to as "incomprehensible," are groups created via algorithmic classification and defined by characteristics and/or definition rules that are not human comprehensible. Concerning incomprehensible characteristics, AI systems frequently use input data containing features that cannot be meaningfully interpreted by a human observer due to their scale (both small and large), volume, complexity, or source. Examples include individual pixels in a picture, mouse movement on a website, web history, or other electronic signals.

Both types of groups are assumed not to correlate with existing protected groups; in cases where such a link can be shown, they can effectively be treated as proxies for protected groups and thus do not pose a novel challenge for anti-discrimination law.

The legal consequences of the widespread creation and usage of such algorithmic groups in automated decision-making remain unclear. Given the rapid global rollout of AI technologies across myriad public and private sectors to make important (and potentially life changing) decisions, it is critical to determine the legal status of algorithmic groups under anti-discrimination law. If these groups do not, or cannot, receive protection under current legal doctrine and practice, it may be necessary to rethink fundamental aspects of the law to keep pace with emerging technologies.

This paper examines the legal status of algorithmic groups in anti-discrimination law in North America and the European Union. It proposes a new theory of harm to close the gap between legal doctrine and emergent forms of algorithmic discrimination. In Section II, I define the novel types of groups that can be created by AI or used for automated decision-making and that thus face risks of bias, inequality, and discrimination.

---

[15] For an overview of commonly used interest categories (including "Winter Activity Enthusiast", "dog owner" and "Heavy Facebook User") see FEDERAL TRADE COMMISSION, *Data brokers: A call for transparency and accountability*, at B-2-B-6 (2014), https://www.ftc.gov/system/files/documents/reports/data-brokers-call-transparency-accountability-report-federal-trade-commission-may-2014/140527databrokerreport.pdf.

[16] Michael Reilly, *Is Facebook targeting advertising at depressed teens?*, MIT TECHNOLOGY REVIEW, https://www.technologyreview.com/s/604307/is-facebook-targeting-ads-at-sad-teens/ (last visited Apr 19, 2019).

[17] Being labelled as a "video gamer" can cause one's Chinese Social Credit Score to drop. See Nicole Kobie, *The complicated truth about China's social credit system*, WIRED UK, 2019, https://www.wired.co.uk/article/china-social-credit-system-explained (last visited Mar 26, 2019).

[18] ARTICLE 29 DATA PROTECTION WORKING PARTY, *Guidelines on Automated individual decision-making and Profiling for the purposes of Regulation 2016/679, 17/EN WP 251rev.01*, 10 and 22 (2018), http://ec.europa.eu/newsroom/article29/document.cfm?doc_id=49826.

[19] The Working Party warns about the possible exploitation of gamers via nudging and online ads see *Id.* at 29.

[20] See Janneke Gerards & Frederik Zuiderveen Borgesius, *Protected grounds and the system of non-discrimination law in the context of algorithmic decision-making and artificial intelligence*, AVAILABLE AT SSRN, 40 and 63 (2020) also argues that socio-economic background could be recognised as a "suspect" category in the future.

In Section III, I then examine case law of the European Court of Justice (ECJ) to determine how protected grounds are determined and how anti-discrimination law applies to algorithmic groups in the first instance. There are two possible ways for a group to qualify for protection: association with a protected group, or as a new protected class. However, the Court has long opposed the idea of adding new groups to the list of protected grounds.[21]

Absent legal reform, algorithmic groups are thus unlikely to be added as a new protected class. AI disrupts the fundamental protection mechanisms of anti-discrimination law. Nonetheless, significant risks of inequality and discrimination remain. It is thus necessary to take a step back to examine the core aims of the law to determine whether algorithmic groups should be added in theory. Future case law might adopt a broader interpretation of the law by appealing to its theoretical foundations.

Across the next three sections, I examine three possible pathways to expand the scope of anti-discrimination law in this fashion, proceeding from practical to abstract. In Section IV, I examine the requirements for groups to receive protection. I propose a novel taxonomy of typical characteristics that make groups worthy of protection focusing on (1) immutability and choice, (2) relevance, arbitrariness, and merit, (3) historical oppression, stigma, and structural disadvantage, and (4) social saliency. I find that algorithmic groups typically do not meet most of these criteria and thus cannot be directly added by appealing to the nature of protected grounds.

Section V moves up a level of abstraction to consider the theoretical underpinnings of the law's account of the moral wrongness of discrimination. In other words, what is discrimination according to the law, and why is it wrong? I examine how the law's foundational account of moral wrongs drives its selection of protected characteristics. Discrimination is framed as a behavior that carries with it an assumption of moral superiority between groups, meaning that the discriminator implicitly or explicitly compares groups and assigns them different worth. For algorithmic groups and decision-makers this mindset cannot be assumed; socially meaningful groups may never be implicitly or explicitly compared in terms of moral worth. Again, this suggests a mismatch between the law's account of the moral wrongness of discrimination and the type of disadvantage experienced by algorithmic groups.

---

[21] Sandra Wachter, *Affinity Profiling and Discrimination by Association in Online Behavioral Advertising*, 35 BERKELEY TECH. LJ 367, 56–60 (2020); Gerards and Zuiderveen Borgesius, *supra* note 20; for the challenges this brings for human rights in general, see Alessandro Mantelero, *AI and Big Data: A blueprint for a human rights, social and ethical impact assessment*, 34 COMPUTER LAW & SECURITY REVIEW 754–772, 765 (2018).

Section VI continues upwards to examine the fundamental doctrines and remit of anti-discrimination law. If the law were to fully achieve its aims, what would society look like? If it can be found that decisions against algorithmic groups contravene these basic doctrines it may be possible to include them. Many scholars agree the law aims to eliminate differences between certain groups, or to create a 'level playing field'. This approach creates a problem for algorithmic groups because it may be impossible to establish the existence of oppression, or to show that certain groups receive better treatment than others. Nonetheless, different theories of anti-discrimination law appear to share some common ground in their commitment to protect basic rights, liberties, freedoms, and access to goods and services. They differ in how best to achieve these aims, but the basic agreement that the law should enable people to make informed and autonomous life choices remains.

While none of the three pathways explored directly bear fruit, I nonetheless argue in Section VII that algorithmic groups should enjoy protection under anti-discrimination law. The core consensus agreed by legal scholars around the law's fundamental aim to protect basic rights, liberties, freedoms, and access to goods and services provides a path forward. These goods are enablers of personal autonomy; scholars value clear and concrete decision criteria because they guarantee predictability, allow control over the process, and enable individuals to make appropriate life choices to achieve their life goals (e.g., to earn good grades to enter university). AI created groups can disrupt this idea and can pose a barrier to pursue life goals. The harms and interests facing these algorithmic groups are the same as those targeted by the law; what is different is *how* they are experienced.

Following this, I propose four key elements of a good decision criteria reflecting the law's fundamental aims: stability, transparency, empirical coherence, and ethical and normative acceptability. AI makes these criteria effectively impossible to meet in practice. Despite this, the harms and interests relevant to algorithmic groups are the same as traditionally protected groups. A clear gap thus exists between the fundamental aims and actual practice of the law. This gap necessitates rethinking and reinterpretation of legal doctrine resulting in the inclusion of those new algorithmic groups under the law.

In Section VIII, I propose a new theory of harm for anti-discrimination law that aims to close this gap between anti-discrimination law and algorithmic groups. I introduce the concept of "artificial immutability" which describes how algorithmic groups act as *de facto immutable* characteristics in practice. I propose five sources of artificial immutability in AI: opacity, vagueness, instability, involuntariness and invisibility, and a lack of social concept. Each of these renders decision-making processes unstable, opaque, empirically incoherent, and ethically and normatively unacceptable. Artificial immutability impedes individuals' autonomy, in addition to their planning ability and certainty about their life choices (e.g., in education and employment). Contrary to most governance strategies, I argue that transparency will not remedy the situation. Shedding light on processes that individuals have no control over will not help them to achieve their life goals. Therefore, a stronger focus needs to be placed on control and autonomy rather than just the transparency of decision processes.

Treating algorithmic groups as immutable characteristics provides a new pathway to bring AI discrimination within the scope of anti-discrimination law. This approach aligns well with prior accounts of the fundamental aims of the law and the normative acceptability of basing decisions on immutable characteristics. I conclude with reflections on how the law can be reformed to account for artificial immutability and highlight a fruitful overlap with prior work in data protection law on the "right to reasonable inferences."

## II. ALGORITHMIC GROUPS AND DISCRIMINATION

There are three types of groups that AI can base its decisions on: (1) existing legally protected groups or proxies thereof, (2) non-protected but human understandable groups, and (3) incomprehensible groups defined by incomprehensible or unknown characteristics and boundaries. The latter two are collectively referred to here as "algorithmic groups."

As mentioned above, non-protected groups are those which, like protected groups (e.g., ethnicity, gender, age), are defined by human comprehensible characteristics and definition rules but do not currently receive protection under anti-discrimination law. Non-protected groups can be created and used by human and algorithmic decision-makers alike. In the eyes of the law these groups do not hold special historical or normative significance. Examples already used for purposes such as advertising and personalized pricing include "dog owners,"[22] "sad teens,"[23] "video gamers,"[24] "single parents,"[25] "gamblers,"[26] and "poor people."[27] AI poses two related risks for non-protected groups: it creates new opportunities to use such groups in decision-making, and it can identify new, significant correlations that factor into important decisions. Both potentially drive inequality. If, for example, a link was identified between video game playing and earning potential, being classified as a "video gamer" could contribute to financial difficulties for group members.[28] It would mean that video gamers lose out on loans, insurances or are excluded from seeing certain (job) ads.

---

[22] COMMISSION, *supra* note 15 at B-2-B-6.
[23] Reilly, *supra* note 16.
[24] Kobie, *supra* note 17.
[25] ARTICLE 29 DATA PROTECTION WORKING PARTY, *supra* note 18 at 10 and 22.
[26] *Id.* at 29.
[27] Gerards and Zuiderveen Borgesius, *supra* note 20 at 40 and 63.

The second type of algorithmic group is referred to as "incomprehensible." These are groups created via algorithmic classification and defined by characteristics and/or definition rules that are not human comprehensible. Concerning incomprehensible characteristics, AI systems frequently use input data containing features that cannot be meaningfully interpreted by a human observer due to their scale (both small and large), volume, complexity, or source. Examples include individual pixels in a picture, mouse movement on a website, web history, or other electronic signals.[29]

Incomprehensible groups need not, however, be defined by incomprehensible characteristics. Rather, the clustering logic or rules that define the boundaries of the groups can also be incomprehensible. Take, for example, friendship clusters defined according to the frequency of word usage in social media postings.[30] While the words themselves and "word frequency" are human comprehensible characteristics, the boundaries or rules created by a classifier which divide users into clusters are not equivalently comprehensible, salient, or otherwise intuitive and sensible. Simply put, they are not the sort of rules individuals or society traditionally use to define groups of people, and they are not self-evidently socially or legally significant.

Before proceeding to an examination of anti-discrimination legal doctrine and jurisprudence, a brief note on the scope and basic provisions of anti-discrimination law is necessary. This paper does not discuss algorithmic discrimination against legally protected groups and proxies, or the issues of preventing and mitigating these which have already been discussed extensively in the literature.[31] Instead, the paper addresses non-protected but human understandable groups as well as incomprehensible groups and the potential to protect them under EU non-discrimination law.³² These groups face a core risk in the context of AI decision-making: proving discrimination will be difficult and arguably impossible without legal reform.

---

[28] For more on this and examples of new group harms, see Wachter, *supra* note 21.

[29] Brent Mittelstadt, Chris Russell & Sandra Wachter, *Explaining Explanations in AI*, PROCEEDINGS OF THE CONFERENCE ON FAIRNESS, ACCOUNTABILITY, AND TRANSPARENCY - FAT* '19 279–288 (2019); Tim Miller, *Explanation in artificial intelligence: Insights from the social sciences*, 267 ARTIFICIAL INTELLIGENCE 1–38 (2019); Jenna Burrell, *How the Machine "Thinks:" Understanding Opacity in Machine Learning Algorithms*, BIG DATA & SOCIETY (2016); CHRISTOPH MOLNAR, INTERPRETABLE MACHINE LEARNING (2020), https://christophm.github.io/interpretable-ml-book/ (last visited Jan 31, 2019).

[30] Kuldeep Singh, Harish Kumar Shakya & Bhaskar Biswas, *Clustering of people in social network based on textual similarity*, 8 PERSPECTIVES IN SCIENCE 570–573 (2016).

[31] For more discussion on traditional forms of discrimination see Sandra Wachter, Brent Mittelstadt & Chris Russell, *Bias preservation in machine learning: the legality of fairness metrics under EU non-discrimination law*, 123 W. VA. L. REV. 735 (2020); *Id.*; Pauline T. Kim, *Auditing Algorithms for Discrimination*, 166 U. PA. L. REV. ONLINE 189 (2017); Solon Barocas & Andrew D. Selbst, *Big data's disparate impact*, 104 CALIFORNIA LAW REVIEW (2016); Tal Zarsky, *The Trouble with Algorithmic Decisions An Analytic Road Map to Examine Efficiency and Fairness in Automated and Opaque Decision Making*, 41 SCIENCE TECHNOLOGY HUMAN VALUES 118–132 (2016); Andrea Romei & Salvatore Ruggieri, *A multidisciplinary survey on*

Both types of groups are particularly vulnerable to AI-driven inequality because of the difficulty that claimants will face in direct and indirect discrimination cases to prove discrimination. Specifically, to bring a case of indirect discrimination a claimant must demonstrate a strong enough connection between the "apparently neutral provision, criterion or practice" creating the harm, a protected attribute, and a significant disadvantage experienced by the claimant (group).³³ Proving a connection between decision criteria and protected attributes is already problematic in human settings due to the evidential threshold that must be met to establish *prima facie* discrimination;³⁴ this difficulty will only increase for non-protected and incomprehensible groups. Assuming the boundaries, members, and defining characteristics of the group can even be identified, which is highly difficult for groups that defy human comprehension, correlations with protected attributes are not self-evident. As the law

---

*discrimination analysis*, 29 THE KNOWLEDGE ENGINEERING REVIEW 582–638 (2014); TOON CALDERS ET AL., DISCRIMINATION AND PRIVACY IN THE INFORMATION SOCIETY: DATA MINING AND PROFILING IN LARGE DATABASES (2013); Philipp Hacker, *Teaching fairness to artificial intelligence: Existing and novel strategies against algorithmic discrimination under EU law*, 55 COMMON MARKET LAW REVIEW 1143–1185 (2018); Philipp Hacker & Emil Wiedemann, *A continuous framework for fairness*, ARXIV PREPRINT ARXIV:1712.07924 (2017); Frederik Zuiderveen Borgesius, *Discrimination, artificial intelligence, and algorithmic decision-making* (2018); Raphaël Gellert et al., *A comparative analysis of anti-discrimination and data protection legislations*, *in* DISCRIMINATION AND PRIVACY IN THE INFORMATION SOCIETY 61–89 (2013); KIM, *supra* note 13; Cynthia Dwork et al., *Fairness through awareness*, *in* PROCEEDINGS OF THE 3RD INNOVATIONS IN THEORETICAL COMPUTER SCIENCE CONFERENCE 214–226 (2012); Moritz Hardt, Eric Price & Nati Srebro, *Equality of opportunity in supervised learning*, *in* ADVANCES IN NEURAL INFORMATION PROCESSING SYSTEMS 3315–3323 (2016); Zachary Lipton, Julian McAuley & Alexandra Chouldechova, *Does mitigating ML's impact disparity require treatment disparity?*, *in* ADVANCES IN NEURAL INFORMATION PROCESSING SYSTEMS 8125–8135 (2018); Ari Ezra Waldman, *Power, Process, and Automated Decision-Making*, 88 FORDHAM L. REV. 613 (2019); A. Koene et al., *UnBias: emancipating users against algorithmic biases for a trusted digital economy* (2018).

³² For more literature on other EU frameworks and AI, such as an assessment of European Convention of Human Rights, see Zuiderveen Borgesius, *supra* note 31; on how the "other status" of Art 14 ECHR could be used to combat discrimination, see Gerards and Zuiderveen Borgesius, *supra* note 20 at 31; see also, SANDRA FREDMAN, DISCRIMINATION LAW 126ff (2011); for an assessment of fundamental EU rights and AI, see J. H. Gerards et al., *Getting the future right: Artificial Intelligence and fundamental rights*, (2020), http://dspace.library.uu.nl/handle/1874/415337 (last visited Feb 7, 2022).

³³ An example would be how bad but apparently neutral working conditions for part-time workers usually affect women more than men. For more on this topic, see Sandra Wachter, Brent Mittelstadt & Chris Russell, *Why fairness cannot be automated: Bridging the gap between EU non-discrimination law and AI*, 41 COMPUTER LAW & SECURITY REVIEW 105567, 36 (2021).

³⁴ Wachter, Mittelstadt, and Russell, *supra* note 33; Wachter, Mittelstadt, and Russell, *supra* note 31.

currently stands, groups cannot enjoy its protection without establishing a link to a protected ground (e.g., ethnicity, gender, age).

Furthermore, strategic grouping or "divide and conquer approaches"[35] can create groups that are diverse enough (e.g., a heterogenous mix of gender and sex) to hide the fact that the decision-making procedure affects one particular protected group (e.g., women) significantly more than others in a similar situation (e.g., men). However, proving a particular disadvantage is necessary to be successful in court. Creating non-traditional groups or audiences (e.g., winter sport enthusiasts) is particularly popular in sectors using AI profiling, such as online behavioral advertising, because it allows advertisers to precisely target demographics.[36]

Algorithmic groups are poised to face increasing inequality as the use of AI profiling and automated decision-making spreads in the public and private sectors. The majority of these technologies, and data science itself, focuses on correlation rather than causation. This challenges the legitimacy of its use.[37] Furthermore, these approaches do not require fully understanding who or what the group consists of to operate. As a result, deployers have little incentive to investigate the members or the attributes that make up algorithmic groups. Unfortunately, even if the incentive structure was changed, the technical limitations in explaining the make-up of the group in human understandable language remains.[38] These incomprehensible groups are not human understandable, and our language has no social concept for them.[39]

This predicted proliferation of risks of bias and inequality for algorithmic groups occurs against the backdrop of a lack of protection under non-discrimination law and the case law of the ECJ.[40] It is therefore urgent to address how to classify, treat and protect new groups that are not

---

[35] Wachter, Mittelstadt, and Russell, *supra* note 33 at 53 and 69.

[36] Wachter, *supra* note 21 at 56–57.

[37] Jonathan Zittrain, *From Technical Debt to Intellectual Debt in AI*, BERKMAN KLEIN CENTER COLLECTION (2019), https://medium.com/berkman-klein-center/from-technical-debt-to-intellectual-debt-in-ai-e05ac56a502c (last visited Oct 2, 2021); PETER GRINDROD, MATHEMATICAL UNDERPINNINGS OF ANALYTICS: THEORY AND APPLICATIONS (2014); Mittelstadt et al., *supra* note 12.

[38] GRINDROD, *supra* note 37; Mittelstadt, *supra* note 3; GROUP PRIVACY: NEW CHALLENGES OF DATA TECHNOLOGIES, (Linnet Taylor, Luciano Floridi, & Bart van der Sloot eds., 1 ed. 2017).

[39] Wachter, *supra* note 21 at 57; Katja De Vries, *Identity, profiling algorithms and a world of ambient intelligence*, 12 ETHICS AND INFORMATION TECHNOLOGY 71–85, 81 (2010); Matthias Leese, *The new profiling: Algorithms, black boxes, and the failure of anti-discriminatory safeguards in the European Union*, 45 SECURITY DIALOGUE 494–511, 501 (2014); Monique Mann & Tobias Matzner, *Challenging algorithmic profiling: The limits of data protection and anti-discrimination in responding to emergent discrimination*, 6 BIG DATA & SOCIETY 2053951719895805, 5–6 (2019) propose to draw on intersectional and post-colonial theories to address emergent discrimination; Jason Yosinski et al., *Understanding neural networks through deep visualization*, ARXIV PREPRINT ARXIV:1506.06579, 2 (2015).

[40] Wachter, Mittelstadt, and Russell, *supra* note 33 at 33, 48 and 67; Wachter, *supra* note 21.

currently protected under the law. It could be argued that the scope of the law ought to be extended.[41]

This might be a very challenging endeavor. As discussed in the following section, the ECJ has historically been reluctant to add new protected groups to the law, even in cases of clear inequality and bias against human-comprehensible groups (e.g., sexual orientation). Algorithmic groups are qualitatively different from prior groups that have sought protection under the law; they are ephemeral[42] or ad hoc groups[43] that may only exist for a brief period of time. Very often individuals will not know they are part of an algorithmic group.[44] These groups can be heterogeneous or randomly assembled. Alternatively, they might be inaccurate (e.g., inferring factually wrong attributes)[45] or have no social meaning at all.[46] Reforming the law to include algorithmic groups will thus clearly be difficult.

### III.    EU NON-DISCRIMINATION CASE LAW AND NEW PROTECTED GROUNDS

Adding new protected grounds to EU non-discrimination law can happen in one of two ways. In general groups are either protected in open, closed, or hybrid lists. In all of these cases, new types of discrimination might find protection under the current scope of existing groups if they are seen as part of or as an extension of the original scope (e.g., obesity as a form of disability). New stand-alone groups however (e.g., poverty) can only be added to open or hybrid lists if the new ground is seen as comparable to the existing ones. Unfortunately, according to the EJC adding algorithmic groups to these lists is unrealistic at the moment.[47]

---

[41] For further reading on ground-breaking and pioneering scholarship examining the protection (and limits) of AI created groups under various laws see GROUP PRIVACY: NEW CHALLENGES OF DATA TECHNOLOGIES, *supra* note 38; Alessandro Mantelero, *From Group Privacy to Collective Privacy: Towards a New Dimension of Privacy and Data Protection in the Big Data Era*, in GROUP PRIVACY 139–158 (2017); Tal Z. Zarsky, *An Analytic Challenge: Discrimination Theory in the Age of Predictive Analytics*, 14 ISJLP 11 (2017); CALDERS ET AL., *supra* note 31; Mittelstadt, *supra* note 3; LEE A. BYGRAVE, DATA PROTECTION LAW: APPROACHING ITS RATIONALE, LOGIC AND LIMITS (2002); Leese, *supra* note 39; Gerards and Zuiderveen Borgesius, *supra* note 20; Alessandro Mantelero, *Personal data for decisional purposes in the age of analytics: From an individual to a collective dimension of data protection*, 32 COMPUTER LAW & SECURITY REVIEW 238–255 (2016); Frederik J. Zuiderveen Borgesius, *Strengthening legal protection against discrimination by algorithms and artificial intelligence*, THE INTERNATIONAL JOURNAL OF HUMAN RIGHTS 1–22 (2020); Edward J. Bloustein, *Group privacy: The right to huddle*, 8 RUTGERS-CAM LJ 219 (1976); Linnet Taylor, *Safety in numbers? Group privacy and big data analytics in the developing world*, in GROUP PRIVACY 13–36 (2017).

[42] Leese, *supra* note 39 at 503; Tal Zarsky, *Transparent predictions* (2013), https://papers.ssrn.com/sol3/papers.cfm?abstract_id=2324240 (last visited Mar 4, 2017).

[43] Mittelstadt, *supra* note 3 at 476.

[44] *Id.*; Mireille Hildebrandt, *Profiling and the Rule of Law*, 1 IDENTITY IN INFORMATION SOCIETY (IDIS), 64 (2009), https://papers.ssrn.com/abstract=1332076 (last visited Jul 31, 2018); Mantelero, *supra* note 41; on what that means for civil and legal action, see Silvia Milano et al., *Epistemic fragmentation poses a threat to the governance of online targeting*, 3 NATURE MACHINE INTELLIGENCE 466–472 (2021); and Mantelero, *supra* note 41 at 245.

[45] Wachter, *supra* note 21.

[46] *Id.* at 57.

This is unfortunate since scholars have long acknowledged that other, novel, and unique types of discrimination such as intersectional discrimination do exist.[48] Even legislators have acknowledged the existence of intersectional discrimination, for example in the Race Directive. Nonetheless the ECJ has not embraced these ideas in its case law yet.[49]

The ECJ stated in Kaltolft that "obesity" cannot be seen as an additional protected attribute to those protected in Directive 2000/78.[50] Protection against discrimination based on obesity can only be granted insofar as it is part of the already existing protections against disability, which in this case the court denied.[51]

A similar assessment was rendered in Coleman,[52] a case centered around discrimination against an employee due to her child's disability, where the court noted that protection under the directive "cannot be extended beyond the discrimination based on the grounds listed exhaustively."[53] The Court made reference to Chacón Navas,[54] where it denied that chronic sickness can be seen as part of the protection against disability. It also explained that chronic sickness cannot be seen as a new protected ground as the listed categories in the Directive are exclusive.

---

[47] For all of this and more detailed discussion of the listing systems and the case law, see Gerards and Zuiderveen Borgesius, *supra* note 20; Gellert et al., *supra* note 31.

[48] Kimberle Crenshaw, *Mapping the margins: Intersectionality, identity politics, and violence against women of color*, 43 STAN. L. REV. 1241 (1990); Devon W. Carbado et al., *Intersectionality: Mapping the movements of a theory*, 10 DU BOIS REVIEW: SOCIAL SCIENCE RESEARCH ON RACE 303–312 (2013); SHREYA ATREY, INTERSECTIONAL DISCRIMINATION (2019).

[49] Recital 14 states that the "aim [is] to eliminate inequalities, and to promote equality between men and women, especially since women are often the victims of multiple discrimination" COUNCIL DIRECTIVE 2000/43/EC OF 29 JUNE 2000 IMPLEMENTING THE PRINCIPLE OF EQUAL TREATMENT BETWEEN PERSONS IRRESPECTIVE OF RACIAL OR ETHNIC ORIGIN, OJ L 180 (2000), http://data.europa.eu/eli/dir/2000/43/oj/eng (last visited Aug 5, 2019).

[50] C-354/13, Fag og Arbejde (FOA), acting on behalf of Karsten Kaltoft, v Kommunernes Landsforening (KL), acting on behalf of the Municipality of Billund, 2014 E.R.R.C. I–246, https://curia.europa.eu/juris/document/document.jsf?docid=160935&doclang=EN (last visited Mar 1, 2022).

[51] JANNEKE GERARDS ET AL., ALGORITHMIC DISCRIMINATION IN EUROPE: CHALLENGES AND OPPORTUNITIES FOR GENDER EQUALITY AND NON DISCRIMINATION LAW 65 (2021), https://data.europa.eu/doi/10.2838/544956 (last visited Oct 13, 2021).

[52] Case C-303/06, S. Coleman v Attridge Law and Steve Law, 2008 E.C.R. I–415, http://curia.europa.eu/juris/document/document.jsf?text=&docid=67793&pageIndex=0&doclang=EN&mode=lst&dir=&occ=first&part=1&cid=6050215 (last visited Mar 26, 2019); for more detailed analysis on the case, see Wachter, *supra* note 10.

[53] CASE C-303/06, *supra* note 52 at para 46.

[54] Case C-13/05, Sonia Chacón Navas v Eures Colectividades SA., 2006 E.C.R. I-06467, para 55 and 56, https://eur-lex.europa.eu/legal-content/en/TXT/?uri=CELEX%3A62005CJ0013 (last visited Oct 16, 2021).

A similar assessment appeared in Parris where a claimant tried to combine the protections against age discrimination with those against sexual orientation.[55] In doing so, they created a new intersectional category. The case centered on a claimant and his civil partner with regards to survival pensions. UK law allowed civil partners to receive a survivor's pension if the civil partnership was formed before the age of 60. This was not possible for the claimant because civil partnerships were not legal at that point.[56] The Court, however, explained that "no new category of discrimination resulting from the combination of more than one of those grounds, such as sexual orientation and age, that may be found to exist where discrimination on the basis of those grounds taken in isolation has not been established."[57]

These cases show that the ECJ has often been very hesitant to grant protection based on grounds that are not listed in the directives. The Court has not been willing to grant protection on grounds such as chronic illness, obesity or in instances where sexuality and age intersect, even though social disadvantage clearly exists. It seems, therefore, rather unrealistic that the court would be open to granting protection to groups such as "video gamers" or "dog owners," let alone groups that defy human comprehension.

The court is thus highly restrictive in opening up the scope of protection of the law. The next possible pathway to consider for the addition of algorithmic groups to anti-discrimination law goes back to the foundations of the law, specifically regarding the theoretical concepts that make a group worthy of protection.

## IV.   WHO IS PROTECTED BY THE LAW AND WHY?

Understanding the logic of the law will help establish if new AI-created groups can be added to the list of protected grounds or if in fact the underlying theory does not apply to these new groups. Treating people differently is, *per se*, not morally or legally problematic. For example, it is legally acceptable to choose romantic partners and friends based on their gender, religious beliefs or sexual orientation.[58] However, it may become a problem if we base certain decision making (e.g., hiring decisions) on the same criteria. So, when does discrimination become legally and morally problematic?

---

[55] The case centerd on an example of intersectionality in which the claimant sued on the basis of the combined factors "age" and "sexual orientation." The measures in isolation did not produce a discriminatory effect. This shows that one protected group needs to meet the threshold of disproportionality. Case C-443/15, David L. Parris v Trinity College Dublin, Higher Education Authority, Department of Public Expenditure and Reform, Department of Education and Skills, 2016 E.C.R. I–897 at para 80, http://curia.europa.eu/juris/document/document.jsf?text=&docid=185565&pageIndex=0&doclang=EN&mode=lst&dir=&occ=first&part=1&cid=7600685 (last visited Aug 14, 2019); for a strong critique see Erica Howard, *EU anti-discrimination law: Has the CJEU stopped moving forward?*, 18 INTERNATIONAL JOURNAL OF DISCRIMINATION AND THE LAW 60–81, at 69 (2018); for an in-depth discussion of this case and on problems with intersectionality in general see Dagmar Schiek, *On Uses, Mis-Uses and Non-Uses of Intersectionality Before the European Court of Justice (ECJ): The ECJ Rulings Parris (C-433/15), Achbita (C-157/15) and Bougnaoui (C-188/15) as a Bermuda Triangle?* (2018).

[56] Wachter, Mittelstadt, and Russell, *supra* note 33 at 20–21.

[57] CASE C-443/15, *supra* note 55 at para 80.

An often-cited approach to the concept of equality comes from Immanuel Kant, whose idea of equality advances that all people are moral equals, deserve equal moral respect, and should only be judged based on their moral choices.[59] Discrimination becomes morally unacceptable when it is based on criteria that are morally irrelevant.[60] But when are criteria morally irrelevant?

The following section will use European and North American scholarship as a starting point to sketch out a taxonomy of morally irrelevant criteria worthy of protection based on the anti-discrimination literature. This is, of course, not the only way to classify these groups and is by no means an exhaustive picture. In fact, there are many other ways to classify protected groups and others have made invaluable contributions to this discourse.[61]

My hope is that the following section will demonstrate that whichever taxonomy one wants to follow, the reader will join me in the conviction that the following common characteristics usually associated with protected groups (or the characteristics described by other scholars) do not easily map on to algorithmic groups. This means that traditional legal theories on the desert of protection cannot be used to argue for the protection of algorithmic groups.

A.    IMMUTABILITY AND CHOICE

One traditional way of thinking about the desert of protection under the law is through the principle of immutability. The underlying logic states that no one ought to be disadvantaged because of an attribute that they had no active role in acquiring, or an attribute they cannot control or change. People should only be judged based on their actions and choices.

---

[58] TARUNABH KHAITAN, A THEORY OF DISCRIMINATION LAW 65 (2015).

[59] Larry Alexander, *What makes wrongful discrimination wrong? Biases, preferences, stereotypes, and proxies*, 141 UNIVERSITY OF PENNSYLVANIA LAW REVIEW 149–219, 200 (1992).

[60] David Feldman, *Civil liberties and human rights in England and Wales*, 53 CAMBRIDGE LAW JOURNAL 610–610, 135 (1994); Cass R. Sunstein, *The anticaste principle*, 92 MICHIGAN LAW REVIEW 2410–2455 (1994); KHAITAN, *supra* note 58.

[61] Fredman explains that the case law describes four rationales (immutability/choice/autonomy; discrete and insular minorities; dignity; and a history of disadvantage) when granting protected status, albeit she explains that every rationale has its limitations and weaknesses. FREDMAN, *supra* note 32 at 130–139; Gerards and Borgesius argue that groups deserve protection if an immutable characteristic forms the basis for decision-making, the decision is based on undue stereotyping or prejudice, the decision is directed at a vulnerable group or a discreet or insular minority or the group has suffered social exclusion and stigmatisation in the past. Gerards and Zuiderveen Borgesius, *supra* note 20 at 36.

Traditionally, race, sex and disability have been seen as immutable attributes that are morally irrelevant.[62]

Whilst it makes sense to use immutability as a starting point, this idea has several limitations. Firstly, not every treatment based on an immutable characteristic is immoral. Secondly, sometimes it is immoral to use characteristics even though we have control or choice over them.

It is clear that making decisions using immutable characteristics such as race will be morally and legally problematic in most cases. Yet, one can also imagine situations where the use of immutable characteristics has an acceptable (or even a desirable) effect. If we were to see sex as an immutable characteristic, legal protection against discrimination based on pregnancy, even though technically discriminatory (i.e., only granted to females), can be seen as socially acceptable (because males cannot be pregnant and are therefore not in need of such protection). In some instances, hiring choices based on sex could in some cases at least be seen as socially acceptable (e.g., not allowing males to work in a rape crisis center for women).[63]

A socially desirable effect might occur in relation to age discrimination. Age is an immutable characteristic, but no one would deem laws against child labor (i.e., having to have a minimum age in order to enter the labor market), as something problematic. Quite the contrary; we would see them as something laudable.[64]

Some categories that were previously seen as immutable are now becoming more fluid. Sex and gender reassignment are good examples of how our conceptions of biology and gender roles are changing. Here, again, it will be unreasonable to expect a person to change their sex or gender in order to avoid discrimination, just because it is possible.[65]

This brings us to the question of choice. Of course, people only ought to be judged based on their choices and actions they have control over, but there are also situations where people should not be pressured into change, just because they can, in order to be successful.

In fact, there are several categories that are technically mutable, but that it would be unreasonable to expect individuals to change. For example, an individual's decision to become pregnant or practice a certain religion can be seen as individual choices, but it would be out of the question to ask an individual to change these characteristics in order to avoid discrimination.[66] At the same time, we might find it acceptable that some religious institutions exclude individuals based on their choices, such as their faith

---

[62] Feldman, *supra* note 60 at 136.
[63] Sophia Moreau, *What is discrimination?*, PHILOSOPHY & PUBLIC AFFAIRS 143–179, 161 (2010); for an overview of global case law on these criteria, see FREDMAN, *supra* note 32 at 131ff.
[64] EVELYN ELLIS & PHILIPPA WATSON, EU ANTI-DISCRIMINATION LAW 3 (2012).
[65] Alexander, *supra* note 59 at 200.
[66] Jessica A. Clarke, *Against immutability*, 125 YALE LJ 2 (2015).

(e.g., by rejecting a non-Catholic job applicant at a Catholic hospice center).[67]

Other traits (e.g., disability) might exist due to prior choices (but not fundamental or important choices), yet they are seen as worthy of protection by the law. Some traits might be the results of past behavior (e.g., liver disease due to alcoholism, or a disability after a car accident) or are chosen out of social commitment (e.g., culturally deaf or taking up caring duties),[68] but we might still want to prevent discrimination based on these traits.[69]

Yet, not every choice and not every immutable characteristic is protected. For example, Shin argues that zodiac signs, singing ability or mathematical skills constitute both immutable characteristics and, for some, even important aspects of their personality, though he believes that decisions made on the basis of these attributes (e.g., in an employment setting) would not constitute discrimination in a legal sense.[70]

To help with this conundrum, Khaitan offers the view that the law's purpose is to "reduce (and ultimately eliminate) pervasive, abiding, and substantial relative disadvantage between certain types of groups (namely groups whose membership is defined by morally irrelevant or valuable personal characteristics)." This view allows us to look at actual group disadvantage and to protect people with both immutable characteristics (e.g., ethnicity) as well as people who have made fundamental choices (e.g., religion). These characteristics ought to be morally irrelevant to our chances of life success.[71]

AI is challenging our traditional understanding of immutability and choice. Zarsky, for example, explains that non-traditional parameters like "sleeping habits" used by AI fall outside the realm of anti-discrimination law because such habits are not central to an individual's self-determination.[72]

The idea of traditional immutability and choice is no longer meaningful because algorithmic groups might not be immutable (e.g., race) but might also not be a fundamental choice (e.g., religion). Non-protected, but human understandable groups, for example dog owners or individuals sharing salient pixels in photos, might not find protection because they are not defined by an immutable characteristic nor a fundamental choice. Despite this, it could be argued that these characteristics should not form the basis of important decisions (see Section VII).

---

[67] Moreau, *supra* note 63 at 161.

[68] Referring to the limits of luck egalitarianism, see Reuben Binns, *Fairness in machine learning: Lessons from political philosophy*, ARXIV PREPRINT ARXIV:1712.03586, 7 (2017) citing; Elizabeth S. Anderson, *What is the Point of Equality?*, 109 ETHICS 287–337 (1999); and Jens Damgaard Thaysen & Andreas Albertsen, *When bad things happen to good people: luck egalitarianism and costly rescues*, 16 POLITICS, PHILOSOPHY & ECONOMICS 93–112 (2017) and their work on "costly rescue."

[69] Moreau, *supra* note 63 at 150 and 156.

[70] Patrick S. Shin, *Is There a Unitary Concept of Discrimination?*, PHILOSOPHICAL FOUNDATIONS OF DISCRIMINATION LAW"(OXFORD UNIVERSITY PRESS, 2013), FORTHCOMING, SUFFOLK UNIVERSITY LAW SCHOOL RESEARCH PAPER, 169 (2013).

[71] KHAITAN, *supra* note 58 at 18 and 50.

[72] Zarsky, *supra* note 41 at 34.

B.     RELEVANCE, ARBITRARINESS, AND MERIT

Irrelevance of the protected trait (e.g., race and gender) in relation to the task at hand (e.g., loan decisions) is another reason why discrimination can be seen as wrong. Race and gender have nothing to do with merit. Using these criteria would amount to arbitrary decision-making. Whilst this logic has great intuitive appeal, on closer inspection these criteria also have problems. Firstly, not every use of protected attributes is arbitrary or irrelevant in the decision-making context. Secondly, even if these criteria were arbitrary or irrelevant, there is no general right that decisions (e.g., hiring) are made based on merit or rational criteria.[73]

Scholars have argued that the irrelevance or arbitrariness of decision criteria is not always a good metric for the immorality or illegality of discrimination. Alexander has stated that for a discriminator the reasons for discrimination can in fact be relevant.[74] Similarly, Khaitan has argued that certain ableist, racist and sexist practices can be relevant for certain jobs.[75] For example, not hiring a woman in fear of future pregnancies or refusing to hire a person of color for the fear of how their customers would react can be seen as relevant or even rational reasons on which to make those decisions. Yet, even in the face of relevance the law condemns such practices.[76]

In addition, many have argued that we do not have a right to certain decision procedures, especially in the private sector. Some have also voiced concerns around this issue.[77] Private autonomy and freedom of contract

---

[73] For a view that non-discriminatory decisions also increase profits, see Sandra Wachter, *How Fair AI Can Make Us Richer*, 7 EUROPEAN DATA PROTECTION LAW REVIEW 367–372 (2021).

[74] Alexander, *supra* note 59 at 151.

[75] KHAITAN, *supra* note 58 at 33; the same view can be found in John Gardner, *On the ground of her sex(uality)*, 18 OXFORD J. LEGAL STUD. 167, 169 (1998) who explains that catering to racist customer preferences "may amount to moral cowardice, or even treachery, depending on the discriminator's actual and professed sympathies and allegiances. But as it stands it is not irrationality."; Alexander, *supra* note 59 at 172; in agreement, see Shin, *supra* note 70 at 168.

[76] Shin, *supra* note 70 at 172.

[77] Sandra Wachter & Brent Mittelstadt, *A Right to Reasonable Inferences: Re-thinking Data Protection Law in the Age of Big Data and AI*, 2 COLUMBIA BUSINESS LAW REVIEW (2019), https://cblr.columbia.edu/wp-content/uploads/2019/07/2_2019.2_Wachter-Mittelstadt.pdf (last visited Sep 25, 2018); Kathleen Creel & Deborah Hellman, *The Algorithmic Leviathan: Arbitrariness, Fairness, and Opportunity in Algorithmic Decision Making Systems*, VIRGINIA PUBLIC LAW AND LEGAL THEORY RESEARCH PAPER, 2–3 (2021) state "that isolated arbitrary decisions are not of moral concern, except when other rights make non-arbitrariness relevant" and that arbitrariness at scale could be morally problematic and this "becomes relevant when arbitrariness occurs at scale because the harm produced by systemic exclusions requires justification."; for examples of harms at scale see O'NEIL,

might be reasons why we do not have, for example, a right to get insurance, to get hired or to get admitted to university.

Many have argued that at least in the private sector[78] random and arbitrary decision-making is not of moral and legal concern, unless it impedes on other rights.[79] In the private sector, deciding to hire somebody based on their zodiac sign, or whether they like puns or purple shoelaces,[80] the first letter of their name,[81] or whether they own a dog is not seen as conflicting with the rights of a loan-, job-, or insurance applicant. Other leave open the possibility that there might be a moral (but not legal) duty to refrain from the use of arbitrary criteria such as eye color or zodiac signs.[82]

Even if merit-based decisions were in fact morally or legally required, scholars like Wasserman[83] and McCrudden call into question whether a neutral yardstick for merit even exists. The concepts of merit and desert are often ill-defined and allow for very different interpretations.[84]

Whilst the scholarship is undecided on whether relevance, arbitrariness and merit play a role in defining the scope of the law, I argue that machines make it clear that relevance should not play a part. The very purpose of AI is to find correlations between data points where humans would see no connection. AI could, for example, see correlations between liking the color green and being creative or liking the color green and being a desirable candidate for a loan, job, or insurance.[85] In other words, AI can make everything relevant. Since data science mainly focuses on correlation and not causation (e.g., if liking the color green actually impacts loan repayment), it can seemingly make any data point or attribute appear relevant. Allowing the relevance criterion to sanction the use of new groups

---

*supra* note 14; Reuben Binns, *Algorithmic Accountability and Public Reason*, PHILOSOPHY & TECHNOLOGY, 549 (2017), http://link.springer.com/10.1007/s13347-017-0263-5 (last visited Apr 24, 2018) who advocates for the standard of "public reason" to applied to AI decisions both done by private and public actors to reign-in arbitrary AI decisions.

[78] It is well established that the public sector has a duty not to act irrationally or arbitrarily, see for example KHAITAN, *supra* note 58 at 33.

[79] See Gardner, *supra* note 75 at 168 who writes "[t]he first is that we owe nobody (or at any rate nobody but ourselves) an across-the-board duty to be rational, so our irrationality as such wrongs nobody (or at any rate nobody but ourselves)."

[80] Creel and Hellman, *supra* note 77 at 2.

[81] KHAITAN, *supra* note 58 at 233–35.

[82] Tarunabh Khaitan & Sandy Steel, *Wrongs, group disadvantage, and the legitimacy of indirect discrimination law*, FOUNDATIONS OF INDIRECT DISCRIMINATION LAW (HART, 2018), 199 (2017).

[83] David Wasserman, *Discrimination, concept of*, 1 ENCYCLOPEDIA OF APPLIED ETHICS 805–814, 807 (1998).

[84] Christopher McCrudden, *Merit principles*, 18 OXFORD JOURNAL OF LEGAL STUDIES 543–579 (1998); more on this topic see, STEPHEN J. MCNAMEE & ROBERT K. MILLER, THE MERITOCRACY MYTH (2009).

[85] This example is inspired by 19th Century (and now debunked) research that ascribed a connection between liking the colour green and being gay, as opposed to liking red or blue and being straight. See JEAN HALLEY, AMY ESHLEMAN & RAMYA MAHADEVAN VIJAYA, SEEING WHITE: AN INTRODUCTION TO WHITE PRIVILEGE AND RACE 20 (2011).

would thus completely disable anti-discrimination law and defeat the purpose of its protection entirely, at least in relation to AI. There would be no restriction of the use of group profiles.

C. HISTORICAL OPPRESSION, STIGMA, AND STRUCTURAL DISADVANTAGE

Another important criterion to be granted protection under anti-discrimination law has to do with historical disadvantage. As the following will show, many scholars have argued that anti-discrimination law ought to prohibit further subordination of already disenfranchised and historically oppressed groups.[86]

Shin, for example, argues that the enumerated groups under anti-discrimination law do not share characteristics with each other. Rather he believes that those groups have in common a history of experiencing significant injustice in a society and that people have a tendency to attribute prejudicial, stereotypical, and bigoted beliefs to them.[87] Shin writes that the unifying principle of discrimination law is to disavow and disallow historic and persistent patterns of unjust inequalities, rather than some unifying trait between the protected groups.[88]

This is an intuitively appealing approach that comes with the caveat that new harms and forms of discrimination might not find protection under the law because they do not have a long history of oppression and subordination.

To open the path for newer forms of oppression, several scholars have provided interesting theories. Drawing on the theories of Katz and Goffman as well as case law (e.g., in the EU and UK), Solanke offers the notion of stigma and power as a criterion that ought to grant certain groups more protection. She offers a list of questions that one should reflect upon to decide whether a particular group is stigmatized and has suffered pervasive and long lasting disadvantage (which is an evolving and culturally-dependent notion).[89] The anti-stigmatization principle offers a departure from the immutability principle and allows new groups to be added to anti-discrimination lists (e.g., overweight people or people with tattoos) but also acknowledges that social change can reintegrate previously stigmatized groups into society (e.g., due to the change of public attitudes towards divorce).[90]

---

[86] See Owen M. Fiss, *Groups and the equal protection clause*, PHILOSOPHY & PUBLIC AFFAIRS 107–177 (1976).

[87] Shin, *supra* note 70 at 169–70.

[88] see *Id.* at 181 Who writes "[t]he best moral explanation for the legal concept of discrimination, in my view, is that it embodies a collection of approaches that together express our society's commitment to identify, disavow, and disallow in our institutional practices the categories of actions that tend to reinforce or resonate with historic and persistent patterns of unjust inequalities."

[89] IYIOLA SOLANKE, DISCRIMINATION AS STIGMA: A THEORY OF ANTI-DISCRIMINATION LAW 162–163 (2016) The checklist includes the following questions: 1. Is the 'mark' arbitrary or does it have some meaning in and of itself? 2. Is the mark used as a social label? 163 3. Does this label have a long history? How embedded is it in society? 4. Can the label be 'wished away'? 5. Is the label used to stereotype those possessing it? 6. Does the stereotype reduce the humanity of those who are its targets? Does it evoke a punitive response? 7. Do these targets have low social power and low interpersonal status? 8. Do these targets suffer discrimination as a result? 9. Do the targets suffer exclusion?

Khaitan also supports the idea that the disadvantage in question does not need to be historically grounded. Rather, he focuses on whether or not disadvantage is likely to appear in the short-term future, which is, of course, more likely to happen if pervasive and substantial disadvantage has happened in the past. Even though he leaves open whether this disadvantage could be material, political or social, he states that the advantage must be "substantial and abiding,"[91] not spurious, more than a minor inconvenience and usually suffered in more than a single aspect of life (e.g., employment, education, housing, health, etc.).[92] The disadvantage that the law wants to rectify does not necessarily need to be traced back to human-created disadvantage, but could also include natural disasters or biological harms.[93] This view allows for the addition of new groups to the scope of protected characteristics (e.g., discrimination based on attractiveness) because there is proven pervasive and abiding disadvantage, as opposed to discrimination based on eye color were such disadvantages do not exist.[94] Gardner too explains that what constitutes harms and injustices will depend on context, institutions, and history.[95]

Moreau also believes that the protected groups have nothing in common other than the fact that the possessor of certain traits (e.g., religion, ethnicity, sexual orientation, sex) should not bear the costs of having them when making certain life choices, for example, when choosing a job or deciding where to shop (e.g., at a candy shop).[96] The reasons for including certain groups might be diverse, yet she links it back to past injustice. She also leaves open the possibility that new harms ought to be prevented as well.[97] However, she explains that subordination ought to be persistent and that it occurs in various social contexts.[98]

---

10. Is their access to key resources blocked?; Eidelson offers a shortened checklist that only consists of three questions: first, whether there is a negative social meaning fixed with that specific trait, second whether this negative meaning is sufficiently entrenched to warrant protection, and third whether the stigma is morally justified. See Benjamin Eidelson, *Solanke, Iyiola. Discrimination as Stigma: A Theory of Anti-discrimination Law. Oxford: Hart, 2017. Pp. 256. $94.00*, 128 ETHICS, 681 (2018).

[90] Of course this leaves open a margin of appreciation and interpretation. For example Solanke describes that people that are smokers and alcoholics ought not to be protected under the law because society does not stigmatise them severely enough whereas war criminals and paedophiles are deserving of their public stigma, see SOLANKE, *supra* note 89 at 36.

[91] Tarunabh Khaitan, *Prelude to a theory of discrimination law*, PHILOSOPHICAL FOUNDATIONS OF DISCRIMINATION LAW, DEBORAH HELLMAN, SOPHIA MOREAU, EDS., OXFORD UNIVERSITY PRESS (FORTHCOMING), OXFORD LEGAL STUDIES RESEARCH PAPER, 149 (2013) and see also at 149 where Khaitan argues that the disadvantage "must be abiding and substantial. It must be abiding in the sense that it must be likely to manifest itself over a certain length of time. It must be substantial in the sense that it should be likely to be more than a minor inconvenience. Substantial disadvantage is usually, although not necessarily, suffered in more than a single, discrete, sphere of human life or activity."

[92] KHAITAN, *supra* note 58 at 35–36; Khaitan, *supra* note 91 at 14–49.

[93] Khaitan, *supra* note 91 at 148; KHAITAN, *supra* note 58 at 30–33.

[94] KHAITAN, *supra* note 58 at 27.

[95] John Gardner, *Liberals and Unlawful Discrimination*, 9 OXFORD J. LEGAL STUD. 1, 22 (1989).

In general, it seems that scholars agree that on some level discrimination law ought to address historical wrongs and that oppression can be culturally dependent. The critical point is that the oppression and disadvantage are real, and not just an abstract possibility or theoretical harm. Harm must actually occur, must be persistent, and must appear in more than one aspect of life.

This is why non-traditional forms of discrimination, for example against left-handed individuals, supporters of a specific sports team,[99] or blue-eyed people,[100] are often seen as not worthy of protection. Whereas discrimination based on low IQ,[101] poverty,[102] HIV status,[103] obesity or attractiveness[104] could be argued to deserve protection of the law.

While it is still debated if groups like people "with lower education, low-income groups, persons characterized by an unhealthy lifestyle, persons with unfavorable genotypes, obese persons, or asylum seekers"[105] are akin to traditionally disadvantaged groups, some regulators have decided to grant protection to some of them.[106] However, the case law also suggests that newer groups experience higher levels of scrutiny than other groups who have a longer history of oppression.[107]

---

[96] Moreau, *supra* note 63 at 150–52, and 156–57.

[97] *Id.* at 179.

[98] SOPHIA MOREAU, FACES OF INEQUALITY: A THEORY OF WRONGFUL DISCRIMINATION 41 (2020).

[99] Zarsky, *supra* note 41 at 16–17.

[100] Alexander, *supra* note 59 at 153; KASPER LIPPERT-RASMUSSEN, BORN FREE AND EQUAL?: A PHILOSOPHICAL INQUIRY INTO THE NATURE OF DISCRIMINATION 33–34 (2014).

[101] KHAITAN, *supra* note 58 at 138.

[102] *Id.*; For more on the connection between technology and poverty, see Linnet Taylor & Hellen Mukiri-Smith, *Human rights, technology and poverty*, RESEARCH HANDBOOK ON HUMAN RIGHTS AND POVERTY (2021), https://www.elgaronline.com/view/edcoll/9781788977500/9781788977500.00049.xml (last visited Feb 7, 2022).

[103] FREDMAN, *supra* note 32 at 110.

[104] In 1992 the City Council of Santa Cruz, California considered to prohibit discrimination based on attractiveness and refusals to conform with conventional standards of dress and appearance, see Richard Arneson, *Discrimination, disparate impact, and theories of justice*, 87 PHILOSOPHICAL FOUNDATIONS OF DISCRIMINATION LAW 105, 101 (2013); in support, see Moreau, *supra* note 63 at 158; See KHAITAN, *supra* note 58 at 27 and 37 who is in favour of protecting appearance, but not eye colour.

[105] Gerards and Zuiderveen Borgesius, *supra* note 20 at 54–55.

[106] Quebec has recognised poverty and "social condition" as a protected ground, see Moreau, *supra* note 63 at 158.

[107] FREDMAN, *supra* note 32 at 113.

As shown, the scholarship seems to agree that a historical component of oppression is at the core of anti-discrimination law. Nonetheless, many scholars also allow protection for new or non-traditional groups under certain circumstances. The criterium to be granted protection under the law, be it as a new or traditional group, is that this group actually (not just theoretically) experiences severe, structural, and abiding disadvantage in various aspects of their lives.

It seems that the underlying idea is this: all people are equal (as defined from a libertarian, dignitarian, or egalitarian viewpoint; see Section VI) and the law steps in when a specific characteristic that is seen as morally and normatively irrelevant is used to systematically suppress groups (e.g., race and in the future maybe poverty or physical appearance). In other words, it is the law's job to keep an eye on whether a specific characteristic falls below a certain acceptable threshold and is used as a vehicle for systemic oppression.

Historical and current exclusion and pervasive disadvantage as well as public stigma work very well in human settings where we have social categories and concepts for groups and individuals. However, with AI this might no longer be the case.

The most obvious tension point is historical disadvantage. Algorithmic groups (e.g., dog owners, video gamers, sad teens, or people sharing a specific pixel or having similar clicking behaviors) are not groups that have suffered pervasive disadvantage in the past.

The previous section has, of course, also shown that historical disadvantage is not necessarily a precondition to deserve protection under the law. Current public stigma and real-life oppression in several aspects of life can move an attribute under the protection of the law.

Yet, it is very doubtful whether those theories of harm can be translated to algorithmic discriminators. The nature and the fabric of stigma might not be the same when algorithms create those groups. Solanke proposed a stigma checklist which is instructive to examine this gap:[108]

1. Is the 'mark' arbitrary or does it have some meaning in and of itself?
2. Is the mark used as a social label?
3. Does this label have a long history? How embedded is it in society?
4. Can the label be 'wished away'?
5. Is the label used to stereotype those possessing it?
6. Does the stereotype reduce the humanity of those who are its targets? Does it evoke a punitive response?

---

[108] SOLANKE, *supra* note 89 at 162–163.

7. Do these targets have low social power and low interpersonal status?
8. Do these targets suffer discrimination as a result?
9. Do the targets suffer exclusion?
10. Is their access to key resources blocked?[109]

These questions suggest that it is highly unlikely that algorithmic groups (e.g., dog owners, pixel groups) would be seen as such a stigmatized group. Put differently, it is doubtful that algorithms can create the same type of stigma as humans.

Can new stigma even arise? Traditional notions of stigma presume that the individual and or society are aware of the attributes of a specific group and attach an assessment or meaning to it (e.g., a stereotype). The relevant question to ask is therefore whether new stigma can occur if individuals or society do not or cannot know that a person possesses a specific attribute. If a person or society at large does not know that they are seen as a video gamer or a person with specific pixels, it is not clear if we are still talking about stigma in the traditional sense. However, even if individuals or society at large were to figure out that they were profiled as video gamers, dog owners or people sharing a specific pixel, it is questionable if group membership is of the same nature as being attractive, a woman, Black, or having a low IQ.

Of course, non-protected, but human understandable groups (like dog owners) can be added to the list if they start to suffer the same oppression as traditional groups. In other words, if new and pervasive harms occur and persist, then non-protected, but human understandable groups (e.g., video gamers) can also find protection under the law.

However, many of these groups are ephemeral. They might only exist for a short period of time. Therefore, not only will new groups not find protection because they have not been historically disadvantaged and are not associated with public stigma, but they might also not find protection because they cannot point to a new emerging and persistent pattern of disadvantage. Individuals might be only temporarily part of that group, then move to another group or even be part of multiple groups.

The ephemeral nature of groups also means that group membership does not travel with individuals through various aspects of their life in the same way that gender identity or ability would. This also means that group membership might not be the basis for oppression in various aspects of life (e.g., in the job market, education). The disadvantage might also not be long lasting (a one-off occurrence) or severe enough (e.g., slightly higher prices). All of these eventualities challenge our traditional way of thinking about persistent disadvantage.

---

[109] *Id.*

Several open questions remain: does the stigma attached to a person need to be known to the person or society? Does the stigma need to be human understandable and have social meaning? And how long does the suffering need to be ongoing to warrant protection under the law? Can this even happen with ephemeral groups? How severely does this new group need to suffer?

D.    SOCIAL SALIENCY: COMMUNAL AND CULTURAL IDENTITY

Another criterion to make groups worthy of protection is "social saliency." The following will show that usually only "salient social groups" are protected under the law.[110] This requirement poses a problem for groups created by AI which may not be socially salient by default.

Wasserman describes social saliency as "groups that can be subject to discrimination…[they] generally have deep social significance: their members are perceived and treated differently in a variety of important respects by the larger society."[111]

For Lippert-Rasmussen, the social saliency criterion is rather strict, requiring that members of the cognate groups as well as the people that are part of the discriminated group are aware of their group membership.[112]

Even though Khaitan also believes in social saliency as a requirement for protection under anti-discrimination law, he describes the term "group" in a generous way, focusing on experienced group disadvantage, rather than whether social saliency stems from personal identity.[113] He convincingly argues that the law purposefully leaves open the question of whether protected groups ought to have a sense of "solidarity, coherence, sense of identity, shared history, language, or culture."[114] He is referring to ageism and ableism where only some of those criteria may apply. In the same vein, he argues that individuals do not need to be aware that they are part of the group.[115]

---

[110] Zarsky, *supra* note 41 at 15; see LIPPERT-RASMUSSEN, *supra* note 100 at 30 who writes that "[a] group is socially salient if perceived membership of it is important to the structure of social interactions across a wide range of social contexts."

[111] Wasserman, *supra* note 83 at 807.

[112] LIPPERT-RASMUSSEN, *supra* note 100 at 34 who writes that "(1) membership is evidenced by a dichotomous distribution of individuals in the relevant and contrasting groups; (2) all individuals are a member of only one group; and (3) it is evident whether or not someone is member of a certain group."

[113] Khaitan, *supra* note 91 at 146; See, MOREAU, *supra* note 98 at 51 who also believes that it is not important if the trait is an important part of the identity. Rather, what matters is whether they are subordinated.

[114] Khaitan, *supra* note 91 at 145; KHAITAN, *supra* note 58 at 30; This view is also supported by the EU case law, see Wachter, *supra* note 21; here the ECJ also did not think that group identity is essential for protection under the law, actual disadvantage was sufficient Case C-83/14, CHEZ Razpredelenie Bulgaria AD v Komisia za zashtita ot diskriminatsi, 2015 E.C.R. I–480, http://curia.europa.eu/juris/document/document.jsf?docid=165912&doclang=EN (last visited Mar 26, 2019).

[115] Khaitan, *supra* note 91 at 145; KHAITAN, *supra* note 58 at 30.

Moreau believes that these groups need to be socially salient, experience subordination and have less relative power than others. Thus, for example, people with bushy eyebrows are not protected under the law.[116] Further she argues that if subordination, other acts that imply inferiority, or governmental failure to offer basic goods for citizens is directed at a random group of people who have nothing in common (i.e., are not socially salient) it would qualify as wrongful action or as a failure to recognize people as equals but would not constitute wrongful discrimination in the legal sense.[117]

McCrudden, on the other hand, believes that, at least in relation to highly prized public goods (e.g., human rights), the focus on group characteristics is not important. Rather, he suggests focusing on whether the public good in question is distributed as equally as possible. The characteristics of the individual person only play a role in so far as they are needed to justify disparity between outcomes.[118]

Eidelson challenges the idea of saliency and the idea that it needs to relate to a historical wrong.[119] He opens up the door of protection against discrimination based on beliefs about climate change, animal cruelty or smoking. Eidelson explains that the decision about whether an act is discriminatory ought not to be determined based on whether a group is socially salient at the time. According to his view, hiring decisions based on hair or eye color ought to be seen as discriminating, even if this group is not socially salient. He leaves open if this is indeed (morally) wrongful discrimination or if it can be justified.[120]

Eidelson also offers an interesting real-life example.[121] In the United States, the Genetic Information Non-discrimination Act (GINA) prohibits employers and health insurers from discriminating against people based on their genetic information. People with genetic predispositions are not individuals that are traditionally stigmatized in our society, yet such predispositions can still lead to (undesirable and illegal) discrimination, because individuals should not bear the costs of the birth lottery.[122]

---

[116] MOREAU, *supra* note 98 at 50.

[117] *Id.* at 160.

[118] C. McCrudden, *Equality and Non-Discrimination'in Feldman*, 11 ENGLISH PUBLIC LAW (OXFORD: OUP, 2004) PARA (2003); Cf ELLIS AND WATSON, *supra* note 64 at 6.

[119] BENJAMIN EIDELSON, DISCRIMINATION AND DISRESPECT 27–30 (2015) He states that "[s]urely what is meant by invoking discrimination in these contexts is not that the policies in question injure people on account of their membership in groups that are important to the structure of social relations. Rather, it is that they objectionably employ these differences between people as grounds for adverse treatment". See also at 169 "[o]ne strength of the disrespect account I have advanced here is that it offers a plausible reconciliation of these different impulses. On the one hand, in principle one can commit the wrong identified here by discriminating against someone because of her race, sexual orientation, weight, favorite color, or nearly anything else. Indeed the central focus is placed not on what one does and why, but on what one fails to do: respect a person by recognizing her for what she is and deliberating accordingly. So it makes no difference what feature of a person one is reacting to in failing to satisfy this obligation—including whether one is biased against a group that is socially salient or historically stigmatized, or whether concordant discrimination is widespread."

[120] *Id.* at 86 He states that "[t]o disrespect someone is to fail to take account of the normative significance of some facet of her moral standing; and it is just not up to a culture to decide what constitutes such a failure."

[121] Eidelson, *supra* note 89 at 682.

The literature generally believes in the need (a) for group members to feel part of a communal and cultural identity and/or (b) for society to have a meaningful social concept for these groups. This shows tensions in relation to AI and social saliency.

Algorithmic groups are very likely to not be socially salient groups. This could be new groups, such as people who like the color green and love ice cream, or incomprehensible groups, such as people that have a certain order of pixels in a picture or similar clicking behavior. These groups do not have meaning in human language and society. For these reasons the idea of communal and cultural identity cannot be applied to algorithmic groups. Individuals cannot, even in the abstract, identify with pixels in a meaningful way. Of course, many scholars challenge the idea that groups need to have a communal identity.

Unfortunately, even those who do not see communal identity or cultural identity as a necessity still believe in a strict social saliency requirement for protection of the law. In fact, the majority of scholars believe in the social saliency requirement. Gender, race or disability are seen as socially salient characteristics. They have meaning for the individual or within society.

AI, on the other hand, is creating groups that are not socially salient in a meaningful way. Society does not have any preconceived notions about these groups or even words for them. In fact, this is one of the main characteristics of algorithmic groups; that they do not map on to any idea of social saliency.[123]

Of course, scholars leave open the idea of having new or non-traditional groups added to the list if they suffer similarly to traditional groups. However, as shown above, even the scholars who focus on actual harm rather than group membership still believe that social saliency is necessary. Therefore, random acts of harm, for example, do not qualify as worthy of protection. This poses an issue with AI because the groups can be very diverse, heterogeneous (or even random), and thus do not qualify for protection under the law.

In the same vein the ECJ ruled that in order to establish grounds of discrimination, one group (however defined) ought to be significantly disadvantaged or experience adversity at a far greater number than others. It is likely that algorithmic groups are not homogeneous enough to not meet that threshold.[124]

---

[122] For a view that this non-salient group ought not to be protected, see LIPPERT-RASMUSSEN, *supra* note 100 at 33.
[123] Zarsky, *supra* note 41 at 32ff.
[124] Wachter, *supra* note 21 at 44–46.

E.　Summary: New groups and the law

The previous section has explored the different theories that explain which groups ought to be protected under the law. While there is significant disagreement in the literature as to which criteria are important, AI seems to disrupt these theories. The idea of immutability and choice (see Section A) is no longer meaningful because algorithmic groups might not be traditionally immutable (e.g., race) but might also not be a fundamental choice (e.g., religion). New groups (e.g., dog owners or having a specific pixel) might not find protection because there are neither immutable nor a fundamental choice.

Algorithmic groups also render the idea of relevant and merit-based decisions meaningless (see Section B). New AI groups render everything relevant and can link it to merit. In fact, that is the core purpose of AI: to find correlations between data points where humans would not think to look. All the sudden, dog ownership could be seen as a relevant decision criterion because it links to job performance.

Similarly, algorithmic groups might not have suffered historical disadvantage and endured oppression (see Section C). These groups might also not be associated with public stigma. The ephemeral nature of the groups, the individual's lack of awareness of their group membership and the sometimes "non-structural harm" or even "trivial harm" make it hard to point to patterns of emerging oppression.

Finally, these groups are not necessarily socially salient (see Section D). These groups are not meaningful in a social and individual sense. They need have no meaning, significance, or reference point in language or culture. This will make it also less likely that AI-defined groups will establish a sense of community or share a common identity which many scholars see as a necessity for protection under the law. Even the scholars who do not believe in a strict identity criterium still see social saliency as a paramount criterion for protection under the law. The majority of the scholarship believes that some sort of social saliency is required.

Even though the previous section has shown that there is considerable debate on whether, and to what extent, all of these or some of these characteristics have to apply, it has become very clear that AI created groups are fundamentally different from traditional groups and none of the traditional characteristics can be easily translated to them.

V.　WHEN AND WHY IS DISCRIMINATION WRONG?

As we have seen in the previous section, algorithmic groups do not map on to the traditional criteria of groups worthy of protection. In order to assess whether algorithmic groups can still be considered worthy of protection, a further level of abstraction is necessary. Since at least some scholars leave open the idea that new groups (e.g., obese people, people with low IQ, people with low socioeconomic status) could be added to the list, it is necessary to assess why we feel that discrimination is wrong in the first place. In other words, a group is seen as worthy of protection if, for whatever reason, society believes that it should not be used as a basis to make decisions about individuals. So, when do we think that it is morally wrong to use specific criteria to make decisions? After reviewing the relevant literature in the following section, I will show that the use of algorithmic groups does not invoke the same wrongfulness that traditional protected groups do. I will show that traditional doctrine on the wrongfulness of discrimination does not see these new groups worthy of protection.

Hellman argued that discrimination is wrong when it is demeaning and when actual power can be used to express this inferiority, regardless of how the discriminated person feels about it.[125] In contrast, Eidelson argued that wrongful discrimination occurs if someone fails to treat a person as an individual and respect their autonomy and their ability to make decisions. This would mean that they are treated with disrespect.[126] Khaitan[127] argues that anti-discrimination law is meant to eradicate "pervasive, abiding and substantial" inequalities between cognate groups because it wrongfully prevents them from living "the good life,"[128] and Moreau[129] claims that discrimination is wrong when it prevents people from realizing their "deliberate freedoms," unfairly subordinates and denies access to basic goods. Alexander[130] believes that discrimination is wrong when a discriminator sees another person as of lesser moral worth. Arneson explains that discrimination is wrongful when it is "driven by unwarranted animus or prejudice."[131] Lippert-Rasmussen[132] focuses on the actual wrong that occurs to the discriminated person to measure wrongful discrimination. Réaume[133] believes that discrimination is wrong when it violates human dignity, Raz[134] and Gardner[135] think that discrimination is wrong when it violates autonomy, freedom, and liberty. MacKinnon believes that discrimination is wrong when it subordinates social groups (e.g., women) and assumes inferiority[136] and Solanke believes discrimination is wrong when it stigmatizes individuals.[137] Arneson also believes that wrongful discrimination does not need to have a racial animus. Profit maximization suffices if the discriminator accepts the collateral damage.[138] He argues that organizing a committee to maintain racial purity in the neighborhood to discourage black people from moving there, not because of their race, but out of fear that the property prices would fall also constitute wrongful discrimination.

---

[125] DEBORAH HELLMAN, WHEN IS DISCRIMINATION WRONG? (2008).

[126] Benjamin Eidelson, *Treating people as individuals* (2013); EIDELSON, *supra* note 119; for a view that every assessment of an individual always includes some sort of generalisation, see FREDERICK SCHAUER, PROFILES, PROBABILITIES, AND STEREOTYPES 68 (2006); and see also, Kasper Lippert-Rasmussen, *"We are all Different": Statistical Discrimination and the Right to be Treated as an Individual*, 15 THE JOURNAL OF ETHICS 47–59 (2011).

[127] KHAITAN, *supra* note 58 at 18.

[128] Khaitan defines the good life as having "at least four basic goods: (i) which will adequately satisfy one's biological needs; (ii) secure one's negative freedom, ie freedom from unjustified interference by others in one's person, projects, possessions, relationships, and affairs; (iii) provide an adequate range of valuable opportunities to choose from; and (iv) are conducive to having an appropriate level of self-respect." *Id.* at 115.

[129] MOREAU, *supra* note 98; Moreau, *supra* note 63 at 147; "Is this person being asked to bear the costs of other people's assumptions about a certain trait of hers, and so not being treated as a being capable of autonomy in her own right?" MOREAU, *supra* note 98 at 109–110 and 157.

[130] Alexander, *supra* note 59 at 159 he calls this "bias."

[131] Richard J. Arneson, *What is wrongful discrimination*, 43 SAN DIEGO L. REV. 775, 775 (2006).

[132] Kasper Lippert-Rasmussen, *The badness of discrimination*, 9 ETHICAL THEORY AND MORAL PRACTICE 167–185, 167 and 174 (2006).

Statistical discrimination has received a lot attention in the literature and is often seen as wrong. For example, stronger security checks for travelers from Muslim-majority countries have been argued to be wrongful regardless of whether these generalizations are accurate.[139] In such cases, Hellman calls for "social solidarity" and to willingly wait longer at the airport.[140] Khaitan also convincingly states that the status quo ought not be entrenched further, even if there was a "statistical justification," because it runs counter to the law's aim of eroding group differences[141]

However, others have argued that generalizations based on statistics are not morally problematic if they are somewhat based on ground truth because they are rational (e.g., not hiring women because they get pregnant) or have even claimed that focusing on the individuals is problematic (e.g., for voting rights) and thus statistical discrimination can be preferable.[142]

All of these views hold one thing in common: discriminatory behaviors carry with them an assumption of moral superiority. In practice this means they demean an individual, consider them as of lower moral value, promote negative stereotypes and prejudice, treat them with disrespect, have a negative impact on their deliberate freedom, autonomy, dignity or life choices or benefits (e.g., job offers or university places), or otherwise constitute stigmatizing or subordinating acts.[143]

---

[133] Denise G. Reaume, *Dignity, equality, and comparison*, PHILOSOPHICAL FOUNDATIONS OF DISCRIMINATION LAW,(OXFORD: OUP).(ESTIMATED-OCTOBER 2013), FORTHCOMING (2013).

[134] JOSEPH RAZ, THE MORALITY OF FREEDOM (1986).

[135] Gardner, *supra* note 95.

[136] CATHARINE A. MACKINNON, FEMINISM UNMODIFIED: DISCOURSES ON LIFE AND LAW (1987).

[137] SOLANKE, *supra* note 89.

[138] Richard Arneson, *Four conceptions of equal opportunity*, 128 THE ECONOMIC JOURNAL F152–F173 (2018).

[139] Binns, *supra* note 68 at 3.

[140] HELLMAN, *supra* note 125; for seminal work on the concept of solidarity, see 33 BARBARA PRAINSACK & ALENA BUYX, SOLIDARITY IN BIOMEDICINE AND BEYOND (2017).

[141] KHAITAN, *supra* note 58 at 188–89 who believes that our statistical proxies that are results of historical injustice cannot justify direct and indirect discrimination (e.g., women are more likely to care for children, a Black cop is less effective than a white cop in a white neighborhood).

[142] Lippert-Rasmussen, *supra* note 126 at 57.

Here again, it becomes clear that these notions make a lot of sense from a human lens. Indeed, this is how humans often see and treat other humans. However, when algorithms are being used it is not fully clear if their grouping invokes the same moral wrong. Zarsky argues that one could at least question whether an automated grouping process and being singled out is demeaning or disrespectful. The same counts as true for the general public that might not feel insulted by such grouping practices not least because they are not aware that such a grouping is happening.[144]

It is of course clear that algorithms can never have the intention to demean people or carry an attitude of moral superiority. This can only be held by the people deploying or developing the algorithm. As mentioned before, very often deployers will not be aware of bias in the dataset or have a clear understanding of what is in the dataset. They will thus not feel that they are superior to whatever they are evaluating. The main purpose is to optimize the process. Of course, not testing for bias could be seen as an indication of purposefully turning a blind eye to the negative effects on disenfranchised people in society.[145] However, it is likely that this result does not stem from a visceral and emotional feeling of superiority but rather "merely" the collateral damage or the negative side effect of the optimization process.

The power relationship between the discriminator and the discriminatee is very often seen as an essential element of anti-discrimination law. One can only subordinate if one holds the power to do so.

Zarsky also calls into question if, for example, a one-off encounter with a bank constitutes a substantial power imbalance between groups.[146] It is not unlikely that the harm is not "significant enough" (e.g., slightly higher prices) and that such matters could be seen as trivial; a conclusion that, whilst criticized by scholars, equality bodies in the EU have arrived at in the past.[147] And finally, Zarsky also argues that the ephemeral nature of algorithmic grouping and the time-limited relationship between the two parties calls into question if a power imbalance exists.[148]

---

[143] Lawrence Blum, *Racial and other asymmetries*, PHILOSOPHICAL FOUNDATIONS OF ANTI-DISCRIMINATION LAW 182–200, 188 (2013).

[144] Zarsky, *supra* note 41 at 26 who also argues that the omission to prevent discrimination (at least in obvious cases) could be seen as demeaning.

[145] Zarksy page 26. It is, of course, also feasible that a deployer or developer could use an algorithm intentionally to hide discriminatory practices. For the purposes of this discussion I set aside this case of "ill intentions" as it would be per se wrong under traditional notions of discrimination and thus not the type of case which poses new harms.

[146] Zarsky, *supra* note 41 at 27 who also argues that this changes if there are no alternatives for the individuals due to market domination.

[147] For more details on EU case law, see Wachter, *supra* note 21 at 41; Zarsky, *supra* note 41 at 30 who also mentioned that even small price variations can have substantial discriminatory symbolic value; in agreement stating that there is no de minimis threshold,

This section has shown that algorithmic groups might not find protection under anti-discrimination law because using these non-traditional groups does not invoke the same moral and legal wrong. Using dog ownership as a criterion to allocate loans does not necessarily mean that a person believes that dog owners or people with specific pixels are of lesser moral worth, that this practice is demeaning, or that having a dog is usually associated with a stereotype or stigma. Finally, an actual power imbalance between the two parties is very often an essential criteria of anti-discrimination law. However, the ephemeral nature and the constant reorganization of these groups might prevent a long lasting and wide-reaching power imbalance between the two parties.

## VI. WHAT IS THE AIM OF ANTI-DISCRIMINATION LAW?

Algorithmic groups thus do not appear to have sufficient similarity with traditional protected groups (see Section IV) and also do not seem to invoke the same moral wrong we would traditionally encounter (see Section V). In a final step, this paper will move one final level of abstraction further away and assess the theoretical underpinnings, purpose, and aims of anti-discrimination law. This section seeks to clarify whether the law's theoretical foundations can be interpreted to grant algorithmic groups protection. In other words, why should we prevent morally and legally wrong discrimination? What is the purpose and aim of the law and why should we comply? If the usage of these new groups runs counter to these fundamental aims and objectives, they could be seen as worthy of protection in the future. The following section will assess what the underlying theory of the aims of anti-discrimination law are and will show that algorithmic groups do not undermine these fundamental aims. This is a barrier to adding them to the realm of protection.

Interestingly enough, theories around anti-discrimination law, for example those explaining why discrimination is morally wrong, have not been discussed at length by (moral) philosophers and legal scholars.[149] One of the reasons might be that it is a relatively new law.[150] Whilst it is easy to have an intuitive response as to why and what type of discrimination is wrong, it is quite hard to pinpoint the criteria that make it so. This, in turn, makes it difficult to describe the aim and purpose of the law. Several

---

see Khaitan and Steel, *supra* note 82; for the view that even trivial life goals (e.g., going dancing or getting tattooed) enjoy protection. KHAITAN, *supra* note 58 at 102–03.

[148] Wachter, *supra* note 21 at 58; Lippert-Rasmussen, *supra* note 126 at 826.

[149] Sophia Moreau, *Discrimination as negligence*, 40 CANADIAN JOURNAL OF PHILOSOPHY 123–149, 124 (2010); EIDELSON, *supra* note 119 at 3. Of course, questions of equality and justice have long been discussed by moral philosophers and legal scholars. The same does not hold true for the theoretical foundations of non-discrimination law.

[150] Khaitan, *supra* note 91 at 161.

scholars believe in the possibility of a unifying theory[151] of anti-discrimination law whilst others are unsure that such a unifying principle exists.[152]

There are three main families of theory that are seen to underpin the aim of the law: egalitarian, libertarian, and dignitarian.[153] Egalitarians focus on equality and treating people equally, whereas libertarians see the aim of the law as being to promote liberty, autonomy, or freedom. Dignitarians believe that discrimination law aims to protect personal dignity, individuality, or personhood.[154] Rather than endorsing a particular approach, the fundamental drivers of anti-discrimination law across these three schools of thought would appear to be freedom, liberty, and the ability and autonomy to pursue life goals. As we will see below (see Section VIII), AI disrupts each of these drivers and therefore poses a risk regardless of one's alignment with a particular theoretical approach.

Across these three schools of thought, most agree that anti-discrimination law is connected to the ideas of equality, fairness, and justice.[155] The idea of equality can at least be traced back to Aristotle,[156] who believed that people are of equal moral worth and that like people deserve like treatment, whereas unlike cases ought to be treated differently.[157] For Aristotle, equality and justice are synonyms.[158]

Ideas of equality also have roots in Judeo-Christian culture[159] and increasingly gained popularity during the Enlightenment in Europe where thinkers like Locke, Paine, and Rousseau were strong advocates of this idea. People are born equal and are of equal moral worth. Of course, at these times, women and slaves were still excluded from this definition.[160]

Equality is nowadays seen as the underpinning principle of all human rights. It is also recognized that equality is the aspirational principle of every modern society, and it has become a guiding principle in the European Union.[161]

---

[151] Shin, *supra* note 70; Khaitan, *supra* note 91; SOLANKE, *supra* note 89.

[152] EIDELSON, *supra* note 119 at 225; Arneson, *supra* note 104 at 102; George A. Rutherglen, *Concrete or Abstract Conceptions of Discrimination?*, VIRGINIA PUBLIC LAW AND LEGAL THEORY RESEARCH PAPER, 129ff (2012); Eidelson, *supra* note 126 at 203.

[153] I will refrain from in-depth analysis of the differences between these three schools of thought. For a comprehensive overview see KHAITAN, *supra* note 58 at 6.

[154] *Id.* at 6–7.

[155] EIDELSON, *supra* note 119 at 2.

[156] G. C. Field, *The Works of Aristotle: Ethica Nicomachea. Translated by WD Ross, MA (Oxford University Press, 1925.)*, 1 PHILOSOPHY 254–255 (1926).

[157] However, Aristotle also famously defended the practice of slavery, see ERNEST BARKER, POLITICS OF ARISTOTLE (1968).

[158] Catherine Barnard, *The Principle of Equality in the Community Context: P, Grant, Kalanke and Marschall: Four Uneasy Bedfellows?*, 57 THE CAMBRIDGE LAW JOURNAL 352–373, 362–363 (1998) citing ARISTOTLE, ETHICA EUDEMIA VII.9.I24ib (W. Ross ed. I925); CF, see Peter Westen, *The empty idea of equality*, HARVARD LAW REVIEW 537–596 (1982).

[159] Bob Hepple, *The aims of equality law*, 61 CURRENT LEGAL PROBLEMS 1, 7 (2008).

[160] Barnard, *supra* note 158 at 362.

[161] *Id.* at 354.

This is also reflected in the scholarship. Fredman convincingly argues that, at least in Europe, anti-discrimination law has the aim of realizing substantive equality with its four dimensions.[162] Hepple also believes that the aim is substantive equality.[163] Collin and Khaitan argue that the aim of both direct and indirect discrimination prohibitions is to eradicate the lingering effects of past discrimination. Direct discrimination is grounded in the idea that everybody is equal and deserves equal treatment whereas prohibitions against indirect discrimination act as a vehicle for substantive equality goals such as valuable opportunities for everybody,[164] even if that means the dominant groups lose out in the short term.[165]

According to Anderson, the aim of equality is to end oppression and domination.[166] Similarly, Hellman argues that anti-discrimination law addresses the moral wrong of the continuing negative effects of historical discriminatory acts. Therefore, she believes that it can be justified to prohibit seemingly neutral provisions (e.g., seniority rules) if they disadvantage historically disenfranchised groups because they are the legacy of past direct discriminatory acts (e.g., the exclusion of women from the workplace).[167] Hellman also argued for the existence of a duty for the decision-maker to vet whether decision criteria are actually a good proxy for the trait of interest.[168] Purposefully using decision criteria that carry this legacy and thereby increasing or magnifying the harm to an individual is morally wrong because it would compound past injustices further. This is true regardless of whether the historical harm was caused by the current decision-maker or not.[169] She convincingly argues that if there are known connections to prior injustice, data cannot purposefully be used, if their use entrenches inequality further,[170] even if that means the loss of profit.[171]

---

[162] Fredman's idea of substantive equality has four dimensions: "the redistributive dimension: breaking the cycle of disadvantage, the recognition dimension: respect and dignity, the transformative dimension: accommodating difference and structural change, the participative dimension: social inclusion and political voice" FREDMAN, *supra* note 32 at 8; Sandra Fredman, *Substantive equality revisited*, 14 INTERNATIONAL JOURNAL OF CONSTITUTIONAL LAW 712–738 (2016).

[163] Hepple, *supra* note 159 at 16.

[164] Hugh Collins & Tarunabh Khaitan, *Indirect discrimination law: Controversies and critical questions*, 26 and 28 (2018).

[165] KHAITAN, *supra* note 58 at 133.

[166] Anderson, *supra* note 68.

[167] Deborah Hellman, *Indirect discrimination and the duty to avoid compounding injustice*, FOUNDATIONS OF INDIRECT DISCRIMINATION LAW, HART PUBLISHING COMPANY 2017–53, 2 (2018).

[168] Deborah Hellman, *Sex, Causation, and Algorithms: How Equal Protection Prohibits Compounding Prior Injustice*, 98 WASH. UL REV. 481 (2020); Reuben Binns, *On the apparent conflict between individual and group fairness*, in PROCEEDINGS OF THE 2020 CONFERENCE ON FAIRNESS, ACCOUNTABILITY, AND TRANSPARENCY 514–524 (2020) who urges to assess if positive label disparities are due to historic discrimination or personal choice.

[169] Hellman, *supra* note 167 at 4 and 10.

Shin argues that the law does "express our society's commitment to identify, disavow, and disallow in our institutional practices the categories of actions that tend to reinforce or resonate with historic and persistent patterns of unjust inequalities."[172] Moreau explains that the aim of the law is to combat the systemic subordination and stigmatization of protected groups and to enable distributive justice created by past injustices.[173]

Moreau also believes that equality is inherently not based on equality, but rather liberty, and thus a comparator is not necessary.[174] This view has been contested by other authors.[175] Moreau advocates that we do not have the rights to certain freedoms because others have it, but because it is owed to all of us independent from each other.[176] These rights are owed regardless of whether the decision-maker had any hand in the prior wrong doing[177] and is something that an individual can expect.[178] This is true even if it meant loss of business or bankruptcy for the decision maker.[179]

By using the idea of stigma, Solanke shifts the focus from the individual and their characteristics to how societies should act and how to hold them complicit for potential harm.[180] Comparisons are drawn between the public health issues of Ebola and the harms that discrimination causes. Society is

---

[170] *Id.* at 9–10.

[171] Hellman, *supra* note 168 who argues that insurance companies ought not to rely on information of prior domestic abuse when granting insurance, regardless of whether this information is a good proxy for future health risks. Information ought not to be used if it is compounding prior injustice regardless of whether the decision-maker had an active role in bringing about that injustice. Hellman's argument is inspired by analogous arguments by medical scientists deeming the re-usage of medical information collected during immoral experiments during the Nazi era as morally unacceptable, even if the scientists did not have a direct hand in the historical injustice.

[172] Shin, *supra* note 70 at 181.

[173] Moreau, *supra* note 63 at 144–45 and 179.

[174] Finding a comparator is often a barrier to invoking rights.

[175] See, KHAITAN, *supra* note 58 at 51–21 who writes that "protected groups suffer relative disadvantage in comparison with cognate groups".

[176] "[...] we have the same amount of freedom as others happen to have, but that each of us has the amount to which he is entitled." Moreau, *supra* note 63 at 151.

[177] Sophia Moreau, *The Moral Seriousness of Indirect Discrimination*, FOUNDATIONS OF INDIRECT DISCRIMINATION LAW, EDITED BY HUGH COLLINS AND TARUNABH KHAITAN 123–48, 127 (2018).

[178] Moreau, *supra* note 149 argues that discrimination ought to be seen as negligence or the failure to do something that a person could reasonably expect.

[179] Moreau, *supra* note 63 at 147 and 167; Moreau, *supra* note 177 at 127"[w]e are more used to thinking that there is something morally troubling about continuing to act in ways that disproportionately disadvantage certain historically stigmatised and underprivileged groups, regardless of the aims or intentions of the agent."; see also Alexander, *supra* note 59 at 170 who believes it is too burdensome to expect decision-makers to fully shoulder these costs, whilst still acknowledging that the problem is valid and needs addressing (e.g., different drinking ages based on gender entrenches problematic stereotypes).

[180] SOLANKE, *supra* note 89 at 102.

responsible for preventing the spread of the "discrimination virus" and for eradicating it.[181]

Contrary to this, it has also been argued that it is unreasonable to ask decision-makers to bear the costs of eliminating bias and injustice. This ought to be the role of reparations, but decision makers should not be forced to close their eyes to the social realities (e.g., accurate statistical discrimination) even if they are a legacy of injustice.[182]

Collins[183] and Barnard[184] offer an additional aim for anti-discrimination: social inclusion. Raz[185] believes that people have a right to freedom and autonomy. Sen[186] and Nussbaum[187] advocate for people's rights to have the opportunity to develop an adequate range of capacities (called the "capabilities approach"). Rawls[188] advocates for equal opportunity for those with comparable skills, abilities and talent and equal willingness to use them. These people should have similar life prospects, regardless of the circumstances they were born into. We therefore ought to eradicate any hindrances that prevent equal opportunity.

Arneson explains that luck egalitarianism is related to this idea.[189] The doctrine states that nobody should be worse off based on things that are not their fault or were not their choice. It has been embraced by several political theorists including Ronald Dworkin, G. A. Cohen, Larry Temkin, Thomas Nagel and John Roemer. This framework comes with the caveat (as Westen points out) that no two people can be completely comparable across all contexts and renders the idea of equality empty.[190]

All these theories have very compelling and strong arguments when applied to historical prejudices and inequalities. They are rooted in the idea that one group has suffered in the past and that the law must take appropriate steps to close the gaps between those groups. These theories have tremendous value and add a clear and strong moral basis for anti-discrimination provisions; however, these accounts are found wanting when algorithms, not humans, take decisions which can create inequality.

---

[181] *Id.* at 105; for an earlier comparison of discrimination to the flu, see VIVIAN GUNN MORRIS & CURTIS L. MORRIS, THE PRICE THEY PAID: DESEGREGATION IN AN AFRICAN AMERICAN COMMUNITY (2002).

[182] Alexander, *supra* note 59 at 163 and 213.

[183] Hugh Collins, *Discrimination, equality and social inclusion*, 66 THE MODERN LAW REVIEW 16–43 (2003).

[184] Catherine Barnard, *The future of equality law: equality and beyond*, THE FUTURE OF LABOUR LAW. LIBER AMICORUM FOR BOB HEPPLE, OXFORD: HART (2004).

[185] RAZ, *supra* note 134.

[186] Amartya Sen, *Capability and well-being73*, 30 THE QUALITY OF LIFE 270–293 (1993).

[187] Martha C. Nussbaum, *Creating capabilities: The human development approach and its implementation*, 24 HYPATIA 211–215 (2009).

[188] JOHN RAWLS, A THEORY OF JUSTICE (2009).

[189] Arneson, *supra* note 138 at 166–167.

[190] Westen, *supra* note 158; for a critique of Rawls and other theories of equality see Arneson, *supra* note 138.

All of these accounts acknowledge that there is one cognate group that is treated better than the other. AI can, of course, advantage and disadvantage certain groups over others, but there is no guarantee that the groupings it uses will make sense to human observers, be socially salient, or otherwise be sensible. Likewise, compared to human decision-makers, AI cannot have comparable conscious biases, prejudices, and stereotypes against people, and cannot intentionally oppress a group.[191]

As a result of both of these attributes, automated decision-making can appear to affected parties to be quite random and arbitrary, and not necessarily linked to a well-defined, human-comprehensible group. Rather, it is more likely that the affected group is rather diverse and unintuitive, at least from the perspective of the affected parties. This also makes the focus on a comparator that receives preferable treatment unsuitable for AI. Finding a suitable comparator is already very hard in the off-line world. With AI, very often there might not be a salient, coherent, or fixed group that receives better or worse treatment. For groups that are not socially salient it will be particularly difficult to assess with whom they should be compared. Identifying comparators is a key mechanism for anti-discrimination law to fulfil its purpose.[192] The disruption of this process by AI suggests that the technology fundamentally undermines the operation of the law, and thus that the law is no longer fit for purpose to mitigate inequality created by AI decision-making.

## VII.  WHY DO ALGORITHMIC GROUPS DESERVE PROTECTION?

The previous sections have shown that it is highly unlikely that algorithmic groups will find protection under anti-discrimination law. Section IV has shown that their composition means they do not align with traditionally protected groups. Section V explained how algorithmic groups do not seem to be subject to the type of 'moral wrong' that the law usually wants to prevent. Finally, Section VI has shown that even the theories underpinning the law's purpose and aims are not violated by the use of non-protected and incomprehensible groups.

The normative and theoretical underpinnings of anti-discrimination law do not accommodate the addition of algorithmic groups to its sphere of protection. They are simply not the type of groups targeted by the law. This naturally begs a question: why should algorithmic groups be protected under the law in the first place? If, as has been shown, they do not fit within the framework of protected grounds and comparators central to anti-discrimination law, why should we change the law to accommodate harms experienced by algorithmic groups?

---

[191] This is, of course, not to say AI is free of biases or prejudices: models can be intentionally designed to advantage certain groups over others, they can learn latent biases and prejudices from their developers and training data, or unintentionally produce unequal outcomes across groups through normal operation. The point here is specifically about conscious and intentional biases and discrimination; intent to discriminate cannot be attributed to narrow AI systems lacking agency or consciousness.

[192] Wachter, Mittelstadt, and Russell, *supra* note 33 at 22–23.

In this section, I propose that these groups are worthy of protection regardless of the history and prior application of anti-discrimination law. While they do not map well onto traditional accounts of its purpose, treatment, and aims, I propose that there is fundamental alignment in terms of harms. Specifically, the harms that automated decision-making can inflict on algorithmic groups is normatively equivalent to the harms that the law wants to prevent. The mode in which the harm is caused and the perpetrator are different, but the harm caused is the same. To ground these claims, I will propose a new theory of harm to underpin anti-discrimination law: the theory of artificial immutability.

Scholars believe that anti-discrimination law fundamentally aims to promote liberty, equality, and/or dignity, and likewise to prevent or redress harms in these terms. Whilst there is much disagreement on the exact interpretation, purpose, and limits of these concepts, most scholars agree that the law ought to guarantee some liberties, freedoms, and access to basic goods to achieve life goals. Basic goods such as to receive an education, to train and pursue a profession, access to healthcare and food, access to financial services (e.g., loans) or housing are seen as areas where the law's power is needed to ensure fair distribution. In other words, the law aims to ensure that people can pursue their life's path on an equal playing field.

This line of thinking is reflected in the ideal of transparency in decision-making. The Civil Rights movement has had great influence on the custom to have more transparent decision criteria in areas such as the workplace, finance, and housing. One of the achievements of this movement was to place more focus on merit and to rein in nepotism, sexism, and racism.[193] At the same time, decision-makers want to maintain some level of autonomy and discretion when making decisions and thereby retain subjectivity.

The best way to think about decision criteria in these areas (e.g., hiring, education, finance) is as some sort of compromise, or as a mix between objective and subjective criteria. On the one hand there are objective criteria (e.g., GPA, income, university degree), and other more subjective criteria (e.g., compatibility with the team, or ability to effectively fulfil the task). Decisions involving both types of criteria reflect a compromise between the competing interests of individual applicants and decision-makers.

Effectively, this is a compromise concerning control and autonomy. Decision-makers want to retain a level of discretion so as to make decisions that best reflect their diverse interests, for example to hire the best performing candidate for a job or give a loan that will likely be repaid. For individuals, control means being able to reliably make decisions about one's life with a view towards future opportunities. Individuals want the ability to make informed choices with a clear sense of the likely impact of their choices on their future opportunities and access to basic liberties, freedoms, goods, and services. While both types of interests are important, anti-discrimination law is best understood as a positive interest aiming to protect the interests of individuals. In practice, the goods the law is designed to protect, including liberty, equality, and dignity, must be balanced against competing interests and goods. Personal control and autonomy are fundamental enablers of these goods.

---

[193] McCrudden, *supra* note 84 at 552.

Following this, I propose four key elements of decision criteria and processes that, if present, enhance personal control and autonomy. To achieve the law's aims of guaranteeing basic liberties, freedoms, and access to goods and services, decision criteria should be (1) stable, (2) transparent, (3) empirically coherent (meaning that they have a proven connection or at least intuitive link to the decision at hand), and (4) normatively and ethically acceptable.

Criteria meeting these requirements enable personal control, meaning that they help individuals reliably plan for the future and access basic goods and services necessary to fulfil their life goals. Transparency is a basic requirement for informed decision-making.[194] To reliably predict and improve the chance of success in a given decision procedure, it is necessary (at a minimum) to be aware of the criteria or rules used to make the decision.[195] (Relative) stability over time is similarly important. Long-term planning is effective only as long as the criteria being planned against continue to be used in the decision procedure. Empirical coherence is important because it allows individuals to observe, infer, or predict links between certain behaviors and decisions and the likelihood of success in a given decision procedure. Similarly, for approaches that value meritorious decision-making, empirical coherence is important to ensure a proven or at least intuitive link exists between the criteria used and the notion of merit underlying the decision procedure. Finally, decision criteria need to comply with social norms and laws, such as anti-discrimination law, to make them normatively and ethically acceptable. Together, these four elements reflect the social contract at the base of anti-discrimination law.

---

[194] There are many philosophical positions that uphold the value of transparency for meaningful choice. For example, J. HABERMAS, THE THEORY OF COMMUNICATIVE ACTION: VOLUME 1: REASON AND THE RATIONALIZATION OF SOCIETY (1984) links transparency to the integrity of public discourse; Patrick Lee Plaisance, *Transparency: An assessment of the Kantian roots of a key element in media ethics practice*, 22 JOURNAL OF MASS MEDIA ETHICS 187–207 (2007) links transparency to the Kantian "principle of humanity" and "supreme principle of the doctrine of virtue," both of which link communication to respect of human dignity. See; IMMANUEL KANT, GROUNDWORK OF THE METAPHYSICS OF MORALS (Mary J. Gregor ed., 1998); Immanuel Kant, *The metaphysics of morals* (1996).

[195] Transparency can of course be achieved through public disclosure of the criteria, but equally through inference from prior assessed cases or other non-universal means.

AI disrupts this social contract. AI relies on unintuitive criteria and non-traditional data. It identifies correlations between a desirable characteristic (e.g., being able to repay a loan, not reoffending, being successful at law school) and seemingly unrelated data (e.g., online activity, shopping behavior, cell phone usage).

Insurance companies in the Netherlands, for example, charge customers that live in flats with certain combinations of numbers and letters (e.g., 4A) higher rates.[196] Creditworthiness can be assessed based on how fast a borrower scrolls through a website and whether or not they use capital letters when filling out forms.[197] If a Facebook user is applying for a loan, but is connected with friends who have previously defaulted on a loan, their application is likely to be rejected.[198] Whether or not an applicant is using a computer at their home or workplace can be a deciding factor for a loan application.[199] Apple users are charged higher prices than PC users[200] and people living in rural areas pay more for online products than people from the city.[201] Job applicants are more likely to succeed if they submit their application via Chrome, rather than a default browser like Internet Explorer or Safari.[202]

Unfortunately, previous sections have shown that this trend of unintuitive, unpredictable, and opaque automated decision-making does not cause legal issues under anti-discrimination law. In fact, it might even be considered 'fairer'.[203] Anti-discrimination law does not guarantee a "right to rational decision-making" that could be used to fight for personal control against unintuitive, unpredictable, and opaque decisions driven by AI. The law is seemingly not violated by the use of arbitrary decision criteria.

---

[196] Gerards and Zuiderveen Borgesius, *supra* note 20 at 2, 15–16.

[197] See Binns, *supra* note 77 at 545 citing Lobosco; see, Katie Lobosco, *Facebook friends could change your credit score*, CNNMONEY (2013), https://money.cnn.com/2013/08/26/technology/social/facebook-credit-score/index.html (last visited Oct 20, 2021).

[198] Lobosco, *supra* note 197.

[199] *Id.*

[200] Larry Dignan, *Mac users pay more than PC users, says Orbitz*, CNET, https://www.cnet.com/tech/services-and-software/mac-users-pay-more-than-pc-users-says-orbitz/ (last visited Oct 20, 2021).

[201] Jennifer Valentino-DeVries, Jeremy Singer-Vine & Ashkan Soltani, *Websites Vary Prices, Deals Based on Users' Information*, WALL STREET JOURNAL, December 24, 2012, https://www.wsj.com/articles/SB10001424127887323777204578189391813881534 (last visited Mar 23, 2019).

[202] See Mark Burdon & Paul Harpur, *Re-conceptualising privacy and discrimination in an age of talent analytics*, 37 UNSWLJ 679, 684 (2014) citing Peck; Don Peck, *They're Watching You at Work*, THE ATLANTIC (2013), https://www.theatlantic.com/magazine/archive/2013/12/theyre-watching-you-at-work/354681/ (last visited Oct 20, 2021).

[203] Basing important decisions on opaque, complex, and unintuitive correlations and data seems intuitively unfair, and yet scholarship would suggest that, at least according to non-discrimination doctrine, the world is now fairer, because decisions are not based on protected groups. Many scholars hold that random decision-making, at least in the private sector, does not cause ethical and legal problems under non-discrimination law; it may, in fact, even be considered fairer than alternatives by not taking protected groups into account. From this perspective, "random" decision-making processes may even be desirable, see Sections IV.B and IV.D.

And yet, algorithmic groups would appear to be vulnerable to significant potential harm measured in terms of personal autonomy and liberties, freedoms, and access to basic goods and services. While AI-driven profiling and decision-making may fail to meet key provisions and requirements of how the law has historically been practiced, it nonetheless would appear to pose threats that match the concerns and values expressed in the law's foundational doctrines. If this is the case, it would appear a new theoretical bridge is required to cross this gap between legal doctrine and application.

## VIII. ARTIFICIAL IMMUTABILITY: A NEW THEORY OF HARM

In this section, I propose a new theory of harm for anti-discrimination law to bridge the gap between the law's aims and its application to AI. The core concept of the proposed new theory of harm is "artificial immutability." AI is preventing individuals from exercising their rights and freedoms and access to important goods and services (the underlying drivers of anti-discrimination law, see Section VI), but in a different way than the law anticipated. The law anticipated that those who hold power will make educational, employment, or loan decisions using criteria that assume the inferior worth of certain groups (e.g., women or people of color), treat them with disrespect, and prevent them from living a free and independent life and further entrench historical disadvantage.

Algorithms, as opposed to humans, do not make decisions based on prejudices or the idea of inferior worth, but in the same way they prevent people from accessing goods and services. They do this by creating groups that effectively act as immutable characteristics that individuals are unable to change and have no control over. As a result, individuals lose the ability to exercise their rights and freedoms, and gain access to goods and services. Therefore, the harm is the same as that originally imagined in anti-discrimination law, only the mode and process of bringing about that harm are different.

According to my proposed theory of artificial immutability, in order for decision criteria to align with the fundamental aims of anti-discrimination law, meaning they enable personal autonomy, planning of one's life, and freedom of choice in access to goods, services, and opportunities, they should be based solely on characteristics within a person's control. To take an example from existing anti-discrimination jurisprudence, decision criteria involving immutable characteristics such as age are *prima facie* unacceptable because they are not within a person's control. Exceptions can be allowed, for example to meet genuine job requirements, but the principled rejection of immutability remains.

AI creates a new type of characteristic which is similarly immutable, or not within a person's control, which I will refer to as "artificial immutability." This concept is distinct from traditional immutability insofar as its rigidity results not from circumstances of birth, genetics, or similarly "natural" sources, but rather from the complexity and inaccessibility of its source. Specifically, AI profiling systems create artificially immutable characteristics which, practically speaking, cannot be controlled by the individual for several reasons. Artificial immutability results from five conditions of how AI creates groups and uses them in decision-making:

1. **Opacity**: Absence of information about key profiling and decision-making processes. There are many respects in which AI profiling and decision-making can be opaque. Individuals may lack information about the purpose of an AI system, the scope and source(s) of data under consideration, the content of the data, the criteria used to make decisions (e.g., thresholds for classifying high- and low-risk cases), and other aspects. Groups formed through profiling processes, or used in decision-making processes, which cannot be accessed by subjects are effectively immutable characteristics due to opacity. Immutability in this respect is variable according to the degree of opacity or absence of information about relevant profiling and decision-making processes. A fully artificially immutable characteristic means the individual cannot know to which group they belong, what defines the group, or its meaning or significance in the decision-making process. Total opacity makes it impossible for individuals to exercise personal autonomy and prepare for a successful outcome.

2. **Vagueness**: Similarly to opacity, vagueness describes a condition of profiling or decision-making processes in which the subject receives inadequate information to make informed choices. For example, a decision-maker could provide information about decision criteria which is abstract or that fails to indicate their relative importance or weights. A bank could explain that friends on Facebook or online behavior are factors in loan or job decisions, but the applicant does not have the ability to assess the significance of these criteria. As a result of receiving vague information, the individual will be unable to reliably assess what 'good' or 'bad' online behavior looks like, which friends are most desirable, and similar aspects. Vagueness makes these decision criteria *de facto* immutable.

3. **Instability**: Many AI systems are not stable, meaning they change over time or produce erroneous or unpredictable behaviors (i.e., edge cases). This is particularly true of AI systems based on machine learning which learn and define decision criteria (e.g., classification rules) over time, in some cases learning 'in the wild' from the environment. Instability of both forms leads to unpredictability, making it effectively impossible for subjects to understand or have meaningful input into the decision-making procedure or to reliably prepare for a positive outcome. Informed choices are not possible if decision-making criteria are instable or shift unpredictably over time.

4. **Involuntariness and invisibility**: Many data points used by AI profiling and decision-making systems are based on involuntary and invisible digital and physiological behaviors that are not self-evidently meaningful. Facial recognition software, for example, is increasingly used for job interviews.[204] These systems can track involuntary muscle responses and other physiological traits like facial expressions, eye movement, retina expansion, respiration, sweat, or heartbeat. Likewise, digital behaviors such as search queries, browsing history, or clicking patterns can be used to group users in advertising audiences. While users may be aware that these data types exist in principle, they need not actively contribute to their collection or usage; rather, they can be captured passively which obfuscates their scope and purpose. Setting aside limited capacities to obscure or control physiological responses (e.g., wearing a mask in public spaces with facial recognition towers), individuals generally lack control over these physiological functions and cannot control what is being measured, understand the significance of the captured data and how it is assessed, and make informed choices about future behaviors (e.g., what is "ideal" eye movement?).

5. **Lack of social concept**: Many incomprehensible algorithmic groups will lack a meaningful reference point in human language. The assembly of pixels in a picture, digital signals from an accelerometer, or clicking patterns are examples that do not have a related social concept; society does not currently distinguish between people or groups in these terms or find these characteristics socially salient. The groups, and the rationale for grouping them, makes sense to an algorithm but not a human. Characteristics that lack a social concept are *de facto* immutable because individuals have no way of changing, impacting, understanding, or attaching meaning to them or decision criteria using them.

To summarize, individuals lack meaningful personal control over artificially immutable characteristics because their inherent meaning and usage in automated decision-making is opaque, vague, and ephemeral. Individuals cannot influence their expression of these characteristics (e.g., eye movement) because they are involuntary and collected invisibly. Unlike traditional immutable characteristics, artificially immutable characteristics defy human understanding and social significance; they have no social concept.

---

[204] *A face-scanning algorithm increasingly decides whether you deserve the job*, WASHINGTON POST, https://www.washingtonpost.com/technology/2019/10/22/ai-hiring-face-scanning-algorithm-increasingly-decides-whether-you-deserve-job/ (last visited Jan 29, 2022).

A.   ARTIFICIAL IMMUTABILITY AND 'GOOD' AUTOMATED DECISIONS

Using artificially immutable characteristics in decision criteria risks failing to meet the four key elements of a good decision criteria outlined above (see Section VII): stability, transparency, empirical coherence, and normative and ethical acceptability.

First, stability is disrupted due to algorithmic decision-making and classification models which learn from data in their environment and are often fragile. Fragility here can be seen in many model behaviors, for example the existence of many edge cases and outliers, error prone or unanticipated performance for certain case types, and the like. As a result, they change over time and may treat cases which look intuitively similar in drastically different ways.[205] New correlations between data points on, for example, whether somebody will be a good worker, will repay a loan, or will do well in law school, are constantly created, updated, and changed. This makes planning ahead challenging.

Second, AI renders many of the benefits of transparency less useful. One of the key reasons people desire transparency in decision-making is so they can plan ahead and adjust their actions to achieve their goals. Transparency can only deliver the desired effect if opaque and vague decision criteria and processes can be described in concrete, well-defined, and stable ways. AI is not deployed to use simple, concrete, clear, and basic decision criteria and logic; this is the type of human decision-making it is meant to augment or replace. The main benefit of AI is to sift through as much data as possible and to learn or create decision rules, and perhaps to learn from its environment. This makes it almost impossible to plan and make informed life choices and to be successful in achieving goals.

Transparency, therefore, loses some of its main benefits because individuals cannot tailor their life choices around that knowledge. More transparency will only partially remedy the problem. Knowing about eye tracing software does not grant power over how one moves their eyes. Knowing about groups that defy human language and are not socially salient does not allow people to put their best foot forward when applying to university.

Anti-discrimination law, and in fact much of legal doctrine, is based on the idea that one should only be judged on actions they have control over. Anti-discrimination law goes to great lengths to restrict decision-making based on immutable characteristics. Unless there is a proven and clear connection between these two the usage will be legally problematic. The heightened scrutiny of the "genuine and determining occupational requirement" when using race in employment decisions is such an example.[206]

---

[205] Singh, Shakya, and Biswas, *supra* note 30.

Transparency that enhances personal control is thus necessary, not simply transparency for its own sake, divorced from social meaning and control. The quality of transparency concerning automated decision criteria, and thus a key determinant of whether these criteria involve artificially immutable characteristics, can be assessed only in terms of its contribution to control over the decision criteria and underlying characteristics.

Thirdly, artificially immutable characteristics and decision criteria often lack empirical coherence, meaning there is no causal or empirically proven relationship between the decision criteria and the decision. As discussed above (see Sections IV.A and VII), usage of immutable characteristics to make legally relevant decisions is rejected in the first instance in anti-discrimination law, but justified exceptions are allowed. For example, it can be socially acceptable to base decisions on immutable characteristics if there is an intuitive or proven connection. Intelligence and school admissions, a height requirement in basketball, or eyesight and pilot licenses are such examples.

Unfortunately, data science and AI are often not concerned with causation or establishing a causal or empirical link between characteristics (e.g., algorithmic groups) and outputs; mere correlation is sufficient to drive decision-making. In other words, decision-makers often have little incentive to empirically prove a causal relationship between decision criteria (e.g., friends on Facebook) and predicted variables (e.g., ability to repay a loan).

In practice, empirical coherence in human decision-making is often as much about intuition as it is about causality and empirical evidence. Take, for example, the well-established presence of gender and race biases in grades.[207] While these biases reflect longstanding social and economic inequalities (among other factors), there is at least a clear intuitive link between grades and the variable(s) being predicted in university admissions (e.g., future academic achievement). In other words, the relationship between good grades and merit in academic terms is at least intuitively (maybe even causally) coherent, whilst fully acknowledging that assessment metrics carry the full legacy of social bias. In a hypothetical fair world where no racist and sexist grading practices existed and where resources were distributed equally, many would argue that using grades as an entry criterion for university would be fair. The same cannot be said for criteria such as dog ownership, zodiac signs, social networks, or social media use, especially if correlation rather than causation suffices to justify its use.[208]

---

[206] See Art 4 in COUNCIL DIRECTIVE 2000/43/EC OF 29 JUNE 2000 IMPLEMENTING THE PRINCIPLE OF EQUAL TREATMENT BETWEEN PERSONS IRRESPECTIVE OF RACIAL OR ETHNIC ORIGIN, *supra* note 49.
[207] HALLEY, ESHLEMAN, AND VIJAYA, *supra* note 85 at 117–139.

While shortcuts are always a problem because they are imperfect proxies for an inherently unpredictable future event (in this case, future academic performance), the usage of proxies that defy even intuitive connections is substantially more problematic in terms of empirical coherence. This is especially true when used by decision-makers or in contexts lacking any incentive to investigate the causal relationship between a proxy and the predicted variable.[209]

The pursuit of mere correlation in AI renders causation disposable. This trend has been criticized by Zittrain, who states that the death of theory and the "intellectual debt" we incur through this practice can be particularly harmful for society.[210] Similarly, Eidelson has argued that an overreliance on statistical evidence could lead to a failure of treating people as individuals, at least in cases where there is reasonable counter evidence available to gather a more accurate assessment of the person, even though this duty has limits too.[211] At a minimum, this tendency can be framed as ethically or normatively problematic from the perspective of anti-discrimination law and personal autonomy. From this perspective we ought to respect people's autonomy and refrain from interfering with their right to make life choices, judge them only based on their actual history, but also fully acknowledge that a person can change in the future.[212]

B.     TRADITIONAL AND ARTIFICIAL IMMUTABILITY

While Sections III through VI have shown how algorithmic groups do not fall within the historical application of anti-discrimination law, the

---

[208] Mittelstadt et al., *supra* note 12.

[209] Wachter and Mittelstadt, *supra* note 77.

[210] Zittrain, *supra* note 37.

[211] EIDELSON, *supra* note 119 at 23; This duty comes with limitations, see Eidelson, *supra* note 126 at 224 where he writes that "there must be limits to what is morally obligatory as a matter of respect. I will leave the question of these limits unresolved—though not without some regret—and rely on the unanalyzed notion of information that is 'reasonably' available to a decision-maker, asking whether it is given 'reasonable' weight."

[212] Eidelson, *supra* note 126 at 204 see also at 216 where he writes that "[a]ccording to this theory, therefore, treating someone as an individual demands two things. First, it means paying reasonable attention to relevant ways in which a person has exercised her autonomy, insofar as these are discernible from the outside, in making herself the person she is. Second, it means recognizing that, because she is an autonomous agent, she is capable of deciding how to act for herself. When we act in accordance with these requirements, we deal with people in a way that respects the role they can play and have played in shaping themselves, rather than treating them as determined by demographic categories or other matters of statistical fate."; For a different view see, Lippert-Rasmussen, *supra* note 126 at 47, 54 and 57 who believes that statistical discrimination does not violate the right to be treated as an individual. In fact, it could be more accurate or beneficial for the individual to be judged based on statistics. Finally, obtaining accurate information is time and cost intensive and is not justified, even if that means to treat some people worse than they deserve.

*48  2022 (forthcoming)  Tulane Law Review*artificial immutability of these groups suggests that their usage is normatively problematic in the context of legal reservations concerning immutability. Many scholars rightly acknowledged that immutability is an undesirable aspect of characteristics used by humans to make decisions. The main critique of positions against immutability has been that positive cases exist (e.g., age discrimination in child labor laws), the fluidity of the concept (e.g., gender identity), and moral objections against requiring a person to change mutable characteristics (e.g., religion, pregnancy).

While these critiques are correct and valid in human settings, immutability does not cause the same concerns when applied to AI. In fact, arguments against immutability can prove very helpful when algorithms are the potential discriminators. Insofar as immutability concerns personal control, arguments concerning the legal status of immutability apply as much to traditional immutable characteristics as they do to AI-defined artificially immutable characteristics. 'Artificial immutability' as proposed here only posits that opacity, vagueness, instability, and a lack of personal control and social meaning render a characteristic *de facto immutable*. The moral or legal significance of this immutability is of secondary importance. Rather, the aim is merely to establish that individuals have no control over the criteria used to make decisions. This leaves open whether the usage of immutable characteristics is morally or legally justifiable.

Questions of justification are highly complex and frequently context dependent. Once decision criteria are known it is possible to assess whether group membership is in fact immutable (e.g., having blue eyes) or *de facto* immutable (e.g., web traffic) or is based on choice (e.g., dog owner, pregnancy) and whether its use is justified. We can then assess whether the use of an immutable characteristic is acceptable, even if members lack control over defining attributes (e.g., high intelligence for university admission). Conversely, if membership depends on choice (e.g., dog owner, pregnancy, religion), we can then assess if people can be expected to change these choices (e.g., wearing specific attire to work, or giving up their dog). Going forward, a context-specific approach to questions of justification, AI, and using immutable characteristics to make important decisions is essential.

### IX.  CONCLUSION

AI systems are now widely used to profile people and make key life decisions about them. While the issues of AI bias and proxy discrimination have been well explored to date,[213] less attention has been given to the harms created by profiling based on groups that do not map on to or correlate with legally protected groups in anti-discrimination law. The way harm is conceptualized (or the theory of harm) in existing anti-discrimination law is a poor fit to capture and protect against the risks that AI creates. Therefore, if we want to ensure our legal frameworks are fit for purpose in the face of emerging technologies, we need a new theory of harm that responds to AI's unprecedented technological capabilities to create harm and inequality.

---

[213] For more discussion on traditional forms of discrimination see Wachter, Mittelstadt, and Russell, *supra* note 31; *Id.*; Barocas and Selbst, *supra* note 31; Zarsky, *supra* note 31; Romei and Ruggieri, *supra* note 31; Kim, *supra* note 6; GERARDS ET AL., *supra* note 51; Zuiderveen Borgesius, *supra* note 31; Janneke Gerards, *The fundamental rights challenges of algorithms*, 37 NETHERLANDS QUARTERLY OF HUMAN RIGHTS 205–209 (2019).

In this paper, I have examined traditional theories of anti-discrimination law from North America and the EU and revealed the difficulties of adding non-traditional algorithmic groups to the law's sphere of protection. I proposed a taxonomy traditionally associated with protected groups and showed that algorithmic groups do not map on to these concepts. I then investigated when and why discrimination is wrong and showed that AI groups do not invoke the same wrongfulness. Finally, I elaborated the aims and goals of anti-discrimination law and explained that they do not fit with the aims for which the law was created. However, I argued that they nonetheless deserve protection. I proposed a new theory of harm based on the concept of "artificial immutability," according to which algorithmic groups should be treated as *de facto* immutable criteria in the context of anti-discrimination law.

A rethinking of emerging harms is necessary to regulate AI decision-making in a reasonable way. This is essential work to ensure people retain autonomy and control over their lives amidst automated decision processes that can be confusing, seemingly arbitrary, and ultimately frustrating. Artificial immutability can serve as a basis for future reform and judicial interpretation of anti-discrimination law to bring the unprecedented harms posed by AI within its scope.

My aim here has not been to establish if and when the use of artificially immutable characteristics is justified, but rather to establish the concept as complementary to traditional notions of immutability in legal scholarship and judicial interpretation. With that said, questions of justifiability will naturally follow any legal adoption of artificial immutability as theoretical grounds to govern equality in AI. To that end, I have argued elsewhere that data driven decision-making necessitates the creation of standards for "reasonable algorithmic decision-making" in the form of justification.[214] I have proposed that individuals should have a "right to reasonable inferences." In other words, individuals should have the right to be reasonably assessed when important life decisions are made about them.

The right requires that AI-driven decision-making needs to be ex ante justified by demonstrating (1) why certain data form a normatively acceptable basis from which to draw inferences; (2) why these inferences are relevant and normatively acceptable for the chosen processing purpose or type of automated decision; and (3) whether the data and methods used to draw the inferences are accurate and statistically reliable.[215] The *ex-ante* justification is bolstered by an additional ex-post mechanism enabling unreasonable inferences, including artificially immutable characteristics, to be challenged.

---

[214] Wachter and Mittelstadt, *supra* note 77 at 2.

If implemented, such a right could provide essential theoretical grounds to assess questions of justifiability for artificially immutable characteristics in anti-discrimination law. Following the law's focus on personal autonomy, strong consideration would need to be given to the question of whether the individual actually has meaningful control over the criteria involved in automated decision-making and is able to change them. Together, the concept of artificial immutability and a right to reasonable inferences can provide a path forward towards meaningful individual control of AI-driven decision-making.

---

[215] *Id.*